\documentclass[aps,prd,nofootinbib,superscriptaddress,twoside,twocolumn,floatfix,a4paper,reprint,preprintnumbers,showkeys]{revtex4-2}

\usepackage{orcidlink,graphicx, color}
\usepackage{booktabs}
\usepackage{comment}
\usepackage[letterspace=-10]{microtype} 

\usepackage{bm, amsmath, amsfonts, amssymb,wasysym,xfrac}
\usepackage{multirow, tabularx, dcolumn, soul}
\usepackage{mathtools, leftidx, braket, slashed, cancel, bigdelim}
\usepackage{blkarray}
\usepackage{pgfplots}
\pgfplotsset{compat=1.18}
\usepackage[figures]{rotating}

\usepackage{tikz}
\usetikzlibrary[plotmarks]
\usepackage{anyfontsize}

\usepackage[utf8]{inputenc} 
\usepackage{hyperref}
\definecolor{OliveGreen}{rgb}{0.33, 0.42, 0.18}
\definecolor{jlab_red}{RGB}{192,39,45}
\definecolor{jlab_orange}{RGB}{249,102,0}
\definecolor{jlab_blue}{RGB}{47,122,121}
\definecolor{jlab_green}{RGB}{65,125,10}
\definecolor{jlab_gray}{gray}{0.6}
\definecolor{magenta}{rgb}{0.5, 0, 0.5}
\definecolor{brown}{rgb}{0.43, 0.17, 0}    % Brown (approx. RGB values for the color)
\definecolor{dorange}{rgb}{0.73, 0.47, 0.05}  % Orange (approx. RGB values)
\definecolor{dyellow}{rgb}{0.49, 0.4, 0.03} 
\newcommand{\ala}{{\it \'a la}}
\newcommand\bef{\begin{figure}}
	\newcommand\eef[1]{\label{fg:#1}\end{figure}}
\newcommand\beq{\begin{equation}}
	\newcommand\eeq[1]{\label{#1}\end{equation}}
\newcommand\beqa{\begin{eqnarray}}
	\newcommand\eeqa[1]{\label{#1}\end{eqnarray}}
\newcommand\bet{\begin{table}}
	\newcommand\eet[1]{\label{tb:#1}\end{table}}
 
\newcommand\eqn[1]{Eq.~(\ref{#1})}

\newcommand\cfex{{\it c.f.}} 
 
\newcommand\lhc{{\it lhc}} 
\newcommand{\rom}[1]{\uppercase\expandafter{\romannumeral #1\relax}}
\definecolor{link_blue}{RGB}{51,102,204}
\definecolor{link_maroon}{RGB}{128,0,0}

\hypersetup{%
	pdftitle={bbud-bsud-tetraquarks},
	pdfsubject={},
	pdfkeywords={},
	colorlinks=true, % Enable colored links
	linkcolor=link_maroon, % Internal document links
	citecolor=link_maroon, % Citation links
	urlcolor=link_maroon, % URLs
	filecolor=black, % File links
}

\def\IMSc{The Institute of Mathematical Sciences, CIT Campus, Chennai, 600113, India}
\def\HBNI{Homi Bhabha National Institute, Training School Complex, Anushaktinagar, Mumbai 400094, India}
\newcommand{\imsc}{\affiliation{\IMSc}}
\newcommand{\hbni}{\affiliation{\HBNI}}
\def\TIFR{Department of Theoretical Physics, Tata Institute of Fundamental Research, Homi Bhabha Road, Colaba, Mumbai 400005, India}
\newcommand{\tifr}{\affiliation{\TIFR}}

\preprint{TIFR/TH/25-6}

\begin{document}
	\title{$bb\bar u\bar d$ and $bs\bar u\bar d$ tetraquarks from lattice QCD using two-meson and diquark-antidiquark variational basis}
	\author{Bhabani Sankar Tripathy~\orcidlink{0000-0001-7759-2778}}
	\email{bhabanist@imsc.res.in}
	\imsc
	\hbni
	
	\author{Nilmani Mathur~\orcidlink{0000-0003-2422-7317}}
	\email{nilmani@theory.tifr.res.in}
	\tifr
	
	\author{M. Padmanath~\orcidlink{0000-0001-6877-7578}}
	\email{padmanath@imsc.res.in}
	\imsc
	\hbni
	\preprint{TIFR/TH/25-6}
	\date{\today}
	\begin{abstract}
		We present a lattice QCD investigation of isoscalar tetraquark systems involving bottom quarks with explicit flavor content $bb\bar{u}\bar{d}$ and $bs\bar{u}\bar{d}$. In the doubly bottom sector, the study focuses on axialvector $J^P=1^+$ quantum numbers, whereas in the $bs\bar{u}\bar{d}$ channel both axial vector $J^P=1^+$ and scalar $J^P=0^+$ quantum numbers are investigated in search of signatures for possible tetraquark bound states. The calculations are performed on four ensembles with dynamical quark fields up to the charm quark generated by the MILC Collaboration, with lattice spacings ranging from approximately 0.058 fm to 0.12 fm, and at different values of the valence light quark mass $m_{u/d}$, corresponding to pseudoscalar meson masses, $M_{ps}=0.5$, 0.6 and 0.7 GeV. The energy eigenvalues in the finite volume are determined by applying a variational procedure to correlation matrices constructed from two-meson interpolating operators and diquark-antidiquark operators. Continuum extrapolated elastic $S$-wave scattering amplitudes of $BB^*$, $KB^*$ and $KB$ are extracted from the ground state eigenenergies following a finite-volume analysis \ala~L\"uscher. The chiral and continuum extrapolated binding energy estimates for the isoscalar axialvector doubly bottom tetraquark $T_{bb}$ from the extracted elastic $BB^*$ $S$-wave scattering amplitudes is found to be $\Delta E_{T_{bb}}(1^+)=-116(^{+30}_{-36})$ MeV. In the $bs\bar{u}\bar{d}$, no statistically significant deviations were observed in the ground state energies from the respective elastic threshold energies, leading to no conclusive evidence for any bound states. 
	\end{abstract}
	\maketitle
	%%%%%%%%%%%%%%%%%%%%%%%%%%%%%%%%%%%%%%%%%%%%%%%%%%%%%%%%%%%%%%%%%%%%%%%%%%%%%%%%%
	
	\section{Introduction}
Despite the early proposals for the existence of complex strongly interacting composite particles \cite{Gell-Mann:1964ewy, Zweig:1964jf} that go beyond a conventional three quark or quark-antiquark picture, only very recently experiments have started unveiling the evidence for such particles. Several recently discovered experimental features point to such novel composite states, generally referred to as exotic hadrons, or typically as XYZ states. A comprehensive review on the present status of the discoveries and their theoretical understanding can be found in Refs. \cite{Chen:2022asf,Brambilla:2019esw,Brambilla:2022ura}. A majority of these fall into the category of heavy tetraquark states, with a few prominent examples being $X(3872)$, $Z_c$, $Z_b$, and $P_c$. A significant recent development in this field is the discovery of the $T_{cc}^+$ tetraquark \cite{LHCb:2021vvq,LHCb:2021auc}, which is unique in requiring four distinct valence quark flavors.
	
	The $T_{cc}^+$ tetraquark falls into the family of doubly heavy tetraquarks, formed out of two heavy quarks and two light antiquarks. Doubly heavy tetraquarks were predicted to be stable under QCD interactions in the heavy-quark mass limit \cite{Carlson:1987hh, Manohar:1992nd, Eichten:2017ffp}. A crucial question is whether the bottom quark is sufficiently heavy for the $bb\bar{u}\bar{d}$ tetraquark to form a bound state below the $BB^*$ two-meson threshold. Several phenomenological investigations utilizing Quark level Models \cite{Ader:1981db,Carlson:1987hh,Silvestre-Brac:1993zem,Brink:1998as, Vijande:2003ki, Janc:2004qn, Vijande:2006jf, Ebert:2007rn, Zhang:2007mu, Karliner:2017qjm, Eichten:2017ffp, Park:2018wjk, Braaten:2020nwp, Tan:2020ldi, Meng:2021agn, Kim:2022mpa, Praszalowicz:2022sqx, Wu:2022gie, Song:2023izj,Meng:2023jqk}, QCD sum rules \cite{Navarra:2007yw, Wang:2017uld, Agaev:2018khe}, effective field theory \cite{Wang:2018atz,Liu:2019stu,Dai:2022ulk,Brambilla:2024thx} and Chromomagnetic Interaction Model \cite{Lee:2009rt,Luo:2017eub,Cheng:2020wxa,Guo:2021yws, Liu:2023vrk} suggest that this is indeed the case. There is a long way to go for the experimental detection of doubly bottom hadrons considering the center of momentum energy it would require to produce a simultaneous pair of bottom quark-antiquark combination. Strategies for the experimental detection of tetraquarks with bottomness-$2$ are discussed in Refs.~\cite{Moinester:1995fk, Ali:2018ifm, Ali:2018xfq}.

	Numerical investigations based on first principles, such as those using Lattice Quantum Chromodynamics (lattice QCD) provide an excellent tool in studying the properties of these exotic states. Early lattice calculations explored the $bb\bar{u}\bar{d}$ tetraquark treating the bottom quarks in the static limit under the Born-Oppenheimer approximation with only light dynamical quark flavors \cite{Bicudo:2012qt,Bicudo:2015vta}. Recently, dynamical simulations involving light and strange (and charm in some studies) quark dynamics have become more common with b-quark evolution more often approximated using a nonrelativistic QCD (NRQCD) evolution on the lattice \cite{Francis:2016hui,Junnarkar:2018twb,Leskovec:2019ioa,Mohanta:2020eed,Hudspith:2023loy,Aoki:2023nzp,Alexandrou:2024iwi, Colquhoun:2024jzh}. Our earlier work also employed the NRQCD framework \cite{Junnarkar:2018twb}. Collectively, these studies provide compelling evidence that the isoscalar axialvector $bb\bar{u}\bar{d}$ channel should host a strong interaction stable bound state tetraquark. In the past few years, lattice investigations have also extended to other heavy tetraquark flavor combinations involving bottom quarks. For instance, systems such as $bc\bar{u}\bar{d}$ have been analyzed in our previous work \cite{Padmanath:2023rdu, Radhakrishnan:2024ihu}, and the $bb\bar{u}\bar{s}$ has been explored in \cite{Leskovec:2019ioa, Hudspith:2023loy}. While there is qualitative agreement on the existence of stable axialvector doubly bottom tetraquarks in the isovector $bb\bar{u}\bar{d}$ and in the isodoublet $bb\bar{u}\bar{s}$ channels, the inferences arrived from different lattice investigations differ in the $bc\bar{u}\bar{d}$ case. A summary and more detailed account of various lattice investigations of doubly heavy tetraquarks can be found in the Refs. \cite{Bicudo:2022cqi,Francis:2024fwf}.

	In this lattice QCD study, we investigate tetraquark systems containing bottom quarks, specifically focusing on the $bb\bar{u}\bar{d}$ tetraquark with quantum numbers $I(J^P)=0(1^+)$ and the $bs\bar{u}\bar{d}$ tetraquark with quantum numbers $I(J^P)=0(1^+)$ and $0(0^+)$. The study of doubly bottom axialvector channel is a substantial extension to our previous investigation \cite{Junnarkar:2018twb}. In this extended work, we utilize new quark-smearing procedures at the sink time slices, as proposed in Ref. \cite{Hudspith:2020tdf}, so as to arrive at conservative energy fit estimates avoiding any potential misidentification of fake plateaus. Note that the main aim of this study is to reliably identify the ground state energies and identify their deviations from noninteracting scenario, if any. In addition, this study involves the use of four different ensembles with two different spatial volumes, facilitating a finite-volume scattering analysis. Such a finite-volume analysis was not performed in our previous work \cite{Junnarkar:2018twb}. Moreover, only a few calculations have attempted such an amplitude analysis \cite{Leskovec:2019ioa,Aoki:2023nzp}.
	
	The study of isoscalar $bs\bar{u}\bar{d}$ channels is a continuation of our previous calculations of a similar tetraquark system with flavor content $bc\bar u\bar d$ \cite{Padmanath:2023rdu, Radhakrishnan:2024ihu}. Only a single lattice QCD study exists addressing this channel \cite{Hudspith:2020tdf}. Several previous studies using QCD sum rules and model calculations have indicated the possibility of bound states for $bs\bar{u}\bar{d}$ with $I = 0$ in these channels. For instance, the quark color delocalization screening model predicts bound states in both quantum channels \cite{Huang:2019otd}. Similarly, a few recent quark model investigations \cite{Chen:2018hts,Tan:2020ldi,Chen:2023syh} support the existence of bound states, although the latter suggests a weaker binding. A QCD sum rule analysis of the scalar bottom-strange channel \cite{Agaev:2019wkk} points to a deeply bound state. However, an early study using a non-chiral model \cite{Zouzou:1986qh} found no evidence for a bound state in the $I(J^P) = 0(1^+)$ channel. The only existing lattice study indicates a weakly bound state for this system and did not include a finite volume analysis \cite{Hudspith:2020tdf}. Naturally, further lattice investigations are required to reach definitive conclusions in these channels.

	The remainder of the draft is structured as follows. 
	In Sec.~\ref{sec:latticesetup}, we summarize the relevant details of the lattice setup utilized. We present various technical details involved in the extraction of finite-volume spectra in Sections \ref{sec:measurements}, \ref{sec:operators}, and \ref{sec:analysis}. The simulated finite-volume eigenenergies are presented in Section \ref{sec:finite_volume}, and the near-threshold scattering amplitudes from them are discussed in Section \ref{sec:ampana}. In section \ref{sec:conclusion}, we summarize the study and reiterate our conclusions. 
	
	%%%%%%%%%%%%%%%%%%%%%%%%%%%%%%%%%%%%%%%%%%%%%%%%%%%%%%%%%%%%%%%%%%%%%%%%%%%%%%%%%%%%%%%%%%%%%%%%
	\section{\label{sec:latticesetup} lattice Setup}
	In this work, we use a set of four lattice QCD ensembles with $N_f=2+1+1$ dynamical flavors respecting the Highly Improved Staggered Quark (HISQ) action generated by the MILC collaboration \cite{MILC:2012znn}. In this setup, we have two different lattice volumes to assess the spatial volume dependence and three different lattice spacings $a$ to investigate the cutoff dependence a potential source of error for any lattice QCD based study of hadrons with heavy quarks, with the finest lattice having $a\sim 0.0582$ fm. The strange and the charm quark masses in the sea are tuned to their respective physical values, whereas the dynamical light quark masses are set to be equal ($m_u=m_d$, isosymmetric) and are chosen to be heavier than their physical values. The gauge field dynamics are tadpole improved and Symanzik improved with coefficients tuned through $\mathcal{O}(\alpha_sa^2, n_f\alpha_sa^2)$ \cite{Follana:2006rc}. We present further details of the ensembles in Table \ref{ensemb_table}.
	
	In the valence sector, we utilize the same setup that was used in our previous investigations \cite{Mathur:2018epb, Junnarkar:2018twb, Mathur:2018rwu,Padmanath:2023rdu, Radhakrishnan:2024ihu}. In this framework, an overlap fermion formulation \cite{Neuberger:1997fp, Neuberger:1998wv} is used to describe the dynamics of valence quark flavors with masses up to the charm quark mass. We compute valence quark propagators for four different quark masses in the quark mass range with representative pseudoscalar meson masses $M_{ps}\sim$ 0.5, 0.6, 0.7, and 3.0 GeV. A landscape plot of the pseudoscalar meson masses studied across the four lattice QCD ensembles can be found in Figure 1 of Ref. \cite{Radhakrishnan:2024ihu}. Note that the chosen $M_{ps}$ values are rather heavier than the physical point and it is more desirable to have further closer-to-physical $M_{ps}$ values. However, the available statistics at lighter $M_{ps}$ values turns out to be insufficient to extract results with reasonable statistical precision. The quark propagators with $M_{ps}\sim$ 0.7 and 3.0 GeV correspond to the physical strange and the charm quarks. The valence strange quark mass was tuned to reproduce the mass of {\it hypothetical} $\eta_s$ meson ($m_{\eta_s}=688.5$ MeV) \cite{Chakraborty:2014aca}, which cannot mix with the physical $\eta$ and $\eta'$ mesons via the self annihilation of $s\bar s$. The valence charm quark mass was tuned by setting the kinetic mass of the charmonium 1S spin average $\{a\overline M_{kin}^{\bar cc} = 0.75 aM_{kin}(J/\psi) + 0.25 aM_{kin}(\eta_c)\}$ on each ensemble studied to the physical charmonium 1S spin average mass (3068 MeV) \cite{El-Khadra:1996wdx}.
	Similar to most other lattice studies of heavy quarkonium, we neglect the OZI suppressed heavy quark self annihilation diagrams in the tuning procedure.
	
	\begin{table}[h]
		\setlength{\tabcolsep}{0.5mm}{
			\centering
			\begin{tabular}{l|c|c|c|c} 
				\hline
				Ensemble & Symbol & Lattice Spacing & Dimension & $M_{ps}^{sea}$ \\
				& & & ($N_s^3 \times N_t$)& \\
				\hline
				$S_1$ & \pmb{\textcolor{red}{\tikz{\pgfsetplotmarksize{1.2ex}\pgfuseplotmark{diamond}}}} & 0.1207(11) & $24^3\times 64$ & 305 \\
				$S_2$ & \pmb{\textcolor{magenta}{\tikz{\pgfsetplotmarksize{1.2ex}\pgfuseplotmark{pentagon}}}} & 0.0888(8) & $32^3\times 96$ & 312 \\
				$S_3$ & \pmb{\textcolor{blue}{\tikz{\pgfsetplotmarksize{1.1ex}\pgfuseplotmark{o}}}} & 0.0582(4) & $48^3\times 144$ & 319 \\
				$L$ & \pmb{\textcolor{OliveGreen}{\pgfsetplotmarksize{1.1ex}\tikz{\pgfuseplotmark{square}}}} & 0.1189(9) & $40^3\times 64$ & 217 \\
				\hline
		\end{tabular}}
		\caption{List of lattice QCD ensembles used in this study. Ensembles $S_1$, $S_2$, and $S_3$ correspond to small volume configurations, while $L$ denotes the large volume ensemble. The lattice spacings quoted are based on the $r_1$ parameter \cite{MILC:2012znn}.}
		\label{ensemb_table}
	\end{table}
	
	Realizing bottom quark dynamics with relativistic action leads to uncontrollable cutoff systematics as $am_b \gg 1$ for most of the lattices available today, where $am_b$ refers to the bare bottom quark mass in lattice units. An alternative to this is the use of NRQCD Hamiltonian \cite{Lepage:1992tx} to describe the bottom quark evolution in the valence sector. The use of NRQCD is justified by the observation of decreasing energy splitting between the bottomonium excitations indicating the largely non-relativistic nature of bottom quarks within, with $v^2 \sim 0.1$, with $v$ being the bottom quark velocity within the bottomonium \cite{Thacker:1990bm,Lepage:1992tx, Bali:1997am}. Naturally, potential models that rely upon the nonrelativistic nature have been successful in describing the low lying spectrum of bottom (as well as charm) hadrons. However, a key drawback of such models is the need for a large set of parameters to achieve sufficient accuracy, complicating calculations and compromising their reliability. 
	
	In the NRQCD framework, the heavy quark mass term is excluded, making quark momentum the highest energy scale in the theory. This exclusion is equivalent to a redefinition of the theory in terms of the inverse heavy quark mass, which allows for the simulation of the bottom quark dynamics on lattices with $a>\frac{1}{m_b}$, mitigating the need for exceedingly fine lattice grids. This introduces two additional complexities: 1) the masses of hadrons studied utilizing this nonrelativistic formulation of quark fields will acquire an additive offset, which needs to be accounted for, 2) the notion of continuum limit for the extracted hadron masses is lost. We address these complexities by working with energy differences rather than extracted hadron masses as discussed later in the draft.

	The NRQCD Hamiltonian considered in this work consists of terms up to $1/(am_b)^2$ and leading order term of $1/(am_b)^3$ and can be expressed as $H=H_0+\delta H$, where the interaction term is given by,
	\begin{align}
		\delta H & = -\frac{c_1}{8(am_b)^3} (D^2)^2 -\frac{c_3}{8(am_b)^2} \vec{\sigma}.\left(\vec{D} \times \vec{E} - \vec{E} \times \vec{D}\right) \nonumber \\
		& +\frac{c_2ig}{8(am_b)^2}\left( \vec{D}.\vec{E} - \vec{E}.\vec{D}\right) -\frac{c_4g}{2am_b} \vec{\sigma}.\vec{B} \nonumber \\
		& +\frac{c_5}{24am_b} D^4 - \frac{c_6}{64(am_b)^2} (D^2)^2, 
	\end{align}
	where $D$ is the finite lattice difference, and $\vec E$ and $\vec B$ are the electric and magnetic components of the gauge field. $c_{1...6}$ are the improvement coefficients that are equal to unity at the tree level. These coefficients $c$ are tuned such that the lattice estimate for the spin averaged kinetic mass of the 1S bottomonia matches the experimental value of 1S bottomonium, \cfex~Refs. \cite{HPQCD:2011qwj,Junnarkar:2019equ,Junnarkar:2018twb,Mathur:2018epb} for details.

	%%%%%%%%%%%%%%%%%%%%%%%%%%%%%%%%%%%%%%%%%%%%%%%%%%%%%%%%%%%%%%%%%%%%%%%%%%%%%%%%%%%%%%%%%%%%%%%%
	%%%%%%%%%%%%%%%%%%%%%%%%%%%%%%%%%%%%%%t%%%%%%%%%%%%%%%%%%%%%%%%%%%%%%%%%%%%%%%%%%%%%%%%%%%%%%%%%%
	%%%%%%%%%%%%%%%%%%%%%%%%%%%%%%%%%%%%%%%%%%%%%%%%%%%%%%%%%%%%%%%%%%%%%%%%%%%%%%%%%%%%%%%%%%%%%%%%
	\section{\label{sec:measurements}Measurements}
	
	The low lying hadron spectrum is extracted from the time dependence of the rest frame two-point current-current correlation matrices in the Euclidean spacetime, whose elements are given by, 
	\beq
	\mathcal{C}_{ij}(t) = \sum_{\mathbf{x}}\left<\Phi_i(\mathbf{x},t)\tilde \Phi_j^{\dagger}(0)\right> = \sum_n \frac{Z_i^n\mathcal{Z}_j^{n\dagger}}{2E^n} e^{-E^nt}.
	\eeq{twoptc}
	Here $\{\Phi_i(\mathbf{x},t)\}$ is a carefully crafted basis of interpolating currents that has the set of quantum numbers of the state we are interested in. The mass or energy of the state can be extracted from the large time behavior of these correlation functions. The strength of operator-state overlap $Z_i^n = \bra{0}\Phi_i\ket{n}$ determines the coupling of the interpolator $\Phi_i$ with the state $n$. 
	
	Like in our previous publications on doubly heavy tetraquarks \cite{Junnarkar:2018twb,Padmanath:2023rdu,Radhakrishnan:2024ihu}, we follow a procedure that involves spatial smearing of the quark fields at the source timeslice to filter out high-momentum modes, which effectively implies the use of a modified quark propagator,
	\begin{equation}
		Q(\bar{x},t;t')=\sum_{\bar{x}'} Q(\bar{x},t;\bar{x}',t'),  
	\end{equation}
	otherwise referred to as the wall-source-smeared quark propagator. Here, primed and nonprimed spacetime indices refer to the source and sink, respectively. This is a non-gauge-covariant smearing procedure and hence is performed on Coulomb gauge fixed gauge configurations as proposed in Ref. \cite{Gattringer:2010zz}. At the sink timeslice, our main results are based on unsmeared quark fields. In a free theory, this setup can be proven to completely eliminate all nonzero momentum modes of the quark fields, whereas in an interacting theory, this is empirically known to efficiently suppress excited state contamination and provide a clean signal for the ground state mass estimates. It is to highlight this difference in quark field smearing at the source and the sink, we utilize a tilde (`~$\tilde{}$~') for the interpolating current at the source timeslice ($t=0$), and use a calligraphic font for the corresponding operator state overlap in \eqn{twoptc}. This difference in the quark smearing procedure at the source and sink timeslice leads to an asymmetric nature of the correlation matrices we compute.

	This asymmetric nature of the correlation matrix elements implies that the correlation matrices are not Hermitian, and hence the eigenvalues are not real. This in turn implies that the approach to the asymptotic value of the effective mass of the eigenvalue correlator, defined as $m_{\mathrm{eff}} = ln(\mathcal{C}(t)/\mathcal{C}(t+1))$, could be oscillatory, leading to potential misidentification of the saturated ground state plateau. One way to avoid such a misidentification is to use a symmetric setup. An obvious symmetric construction with the wall smearing setup is a wall source to wall sink procedure. However, such a procedure is known to be plagued by statistical uncertainties that can significantly subside the signal. Alternatively, one could try to make a comparative study of the asymptotic values with a setup that approaches the symmetric nature. Such a procedure was proposed and referred to as box-sink in Ref. \cite{Hudspith:2020tdf}, and was utilized in our calculations for $\mathcal{D}_{6b}$ \cite{Mathur:2022ovu} and $T_{bc}$ \cite{Padmanath:2023rdu} to verify the asymptotic values of the respective finite volume ground state eigenenergies. This involves the use of a modified quark propagator,
	\begin{equation}
		\tilde{Q}(\bar{x},t;t')=\sum_{|\bar{y}-\bar{x}|<R} Q(\bar{y},t;,t'),
	\end{equation}
	in the computation of correlation functions, where $R$ is the radius of the sphere centered at $\bar{x}$. We utilize the same procedure in this calculation to assess the robustness of our estimate for $T_{bb}$ binding energy.
	
	%%%%%%%%%%%%%%%%%%%%%%%%%%%%%%%%%%%%%%%%%%%%%%%%%%%%%%%%%%%%%%%%%%
	%%%%%%%%%%%%%%%%%%%%%%%%%%%%%%%%%%%%%%%%%%%%%%%%%%%%%%%%%%%%%%%%%%
	\section{\label{sec:operators}Interpolating currents}
	There are three quantum channels of interest in this study. First one is isoscalar axialvector doubly bottom tetraquarks $T_{bb}$ with flavor content $bb\bar u\bar d$, where we reconsider the finite-volume spectrum from our work reported in Ref. \cite{Junnarkar:2018twb} and extend it to involve a finite volume analysis. A spin-zero $bb\bar{u}\bar{d}$ tetraquark is disallowed by symmetries. The remaining two channels are that of the isoscalar bottom-strange tetraquarks $T_{bs}$ with axialvector and scalar quantum numbers and flavor content $bs\bar u\bar d$, analogous to our work on $T_{bc}$ reported in Refs. \cite{Padmanath:2023rdu,Radhakrishnan:2024ihu}. For studying any four-quark state, color singlet interpolating currents can be built out of two quarks and two antiquarks in two forms: Meson-meson type and diquark-antidiquark type, wherein the mesons are separately color singlet projected in the former. We utilize both forms of these operators in our analysis to extract the finite volume ground state energy in the respective channel. 
	
	It is phenomenologically expected that doubly heavy tetraquark system could be deeply bound in the heavy quark limit \cite{Francis:2016hui,Czarnecki:2017vco}. Such a deeply bound system could be a dominantly compact object, which suggests the need of local diquark-antidiquark operators in the basis to probe the respective Fock components, while investigating doubly heavy tetraquark system. Empirically it is also observed that such operators have rich overlap with the ground states in doubly bottom tetraquarks  \cite{Hudspith:2020tdf,Meinel:2022lzo,Hudspith:2023loy,Leskovec:2019ioa,Junnarkar:2018twb,Bicudo:2017szl,Francis:2016hui,Bicudo:2015kna}. In this work, the diquark-antidiquark interpolators are realized with all (anti)quark fields jointly projected to zero momentum. In the color space, one could consider a triplet $3_c$ or antisextet $\bar{6}_c$ representations of $SU(3)_c$ for antidiquarks, together with conjugates of these for the diquarks to form total color singlet tetraquark interpolators. It is theoretically argued and empirically known that the isoscalar scalar light antidiquark configuration $[\bar u\bar d]$ in the $3_c$ has the lowest energy compared to other antidiquark configurations~\cite{Francis:2021vrr,Jaffe:2005zz,Bicudo:2015vta}. Hence we consider only diquark-antidiquark operators built out of $\bar{3}_c-3_c$ configurations in our analysis.  
	
	In this study, we consider wall-smeared quark sources and unsmeared quark sinks in the computation of the correlation functions. The wall-smearing at the source does not allow separate momentum projection of the single meson components in the two meson operators. Whereas at the sink, a separate momentum projection of individual meson components, to build bilocal two-meson interpolators as in Ref. \cite{Padmanath:2022cvl,Alexandrou:2023cqg,Alexandrou:2024iwi,Whyte:2024ihh}, amounts to a nested double summation over the spatial volume in our setup. This procedure being prohibitively expensive particularly for larger lattices, we resort to our conventional procedure of using local two-meson interpolators at the sink. Alternatively, one could perform quark-antiquark contractions in each meson with distinct $\mathcal{Z}_2$ noises; however, this leads to correlation matrices that largely deviate from a Hermitian correlation matrix setup. Hence, realization of bilocal two-meson interpolators remains to be beyond the scope of this work, and we hope to utilize a better suited setup to consider bilocal interpolators in future studies.  In the following, we list the operator basis we have used in all the three quantum channels that we study.

	\ul{\it $bb\bar{u}\bar{d}$ with $I(J^P)=0(1^+)$}:
	The lowest relevant two body scattering channel in this case corresponds to the $BB^*$, whereas the lowest inelastic scattering channel is $B^*B^*$. Hence, for the isoscalar axialvector $bb\bar{u}\bar{d}$ tetraquark, we use the same set of interpolators as was utilized in Ref. \cite{Junnarkar:2018twb}. The lowest three body scattering channel $BB\pi$ is sufficiently high, and any associated left hand nonanalyticities are beyond scope of this work. We consider two operators, one related to the $BB^*$ channel and a second one of the diquark-antidiquark form, as given below. We ignored any operators related to $B^*B^*$ and assume effects from this inelastic channel are negligible compared to our statistical uncertainties. More recently, it was argued by the authors of Ref. \cite{Bicudo:2016ooe,Aoki:2023nzp}, that binding energy of $T_{bb}$ could be influenced by the inelastic $B^*B^*$ channel, such that it decreases in magnitude as observed from a potential framework of extracting scattering amplitude. In this analysis following a scattering analysis, \ala~L\"uscher, using ground state energies from a wall-source setup, we ignore the effects of the $B^*B^*$ channel, and hence omit the related operators from the analysis. Below, we present the operator basis used in this channel. 
	\begin{align}
		\Phi_{\mathcal{M}_{BB^*}}(x) &=&& \left[ \bar{u}(x)\gamma_i b(x)\right] \left[ \bar{d}(x)\gamma_5 b(x)\right]& \nonumber \\
		&& -& \left[ \bar{u}(x)\gamma_5 b(x)\right] \left[ \bar{d}(x)\gamma_i b(x)\right], &\nonumber \\
		\Phi_\mathcal{D}(x) &=&& \big[ \left(\bar{u}(x)^T C\gamma_5\bar{d}(x) - \bar{d}(x)^T C\gamma_5\bar{u}(x)\right) &\nonumber \\
		&&\times& (b^T(x)C\gamma_i b(x))\big].
		\label{eq:bb1}
	\end{align}
	where $C=i\gamma_y \gamma_t$ being the charge conjugation matrix. The quantity inside the square brackets is realized as color singlet objects. We consider the diquark-antidiquark operator with attractive color-antitriplet diquarks \cite{Jaffe:2005zz}, in which the light anti-diquark forms an isoscalar scalar configuration, whereas the trivially flavor symmetric heavy diquark carries axialvector quantum numbers.
	
	\ul{\it $bs\bar{u}\bar{d}$ with $I(J^P)=0(1^+)$}: 
	In this case, our focus is on $S$-wave scattering of $K\bar{B}^*$ scattering in the rest frame leading to infinite volume quantum numbers $I(J^P)=0(1^+)$, which reduces to the $T_1^+$ finite-volume irrep. We use an operator basis similar to what was utilized in Ref. \cite{Padmanath:2023rdu} for the study of isoscalar axialvector $bc\bar u\bar d$ tetraquarks, with the replacement of charm quark with a strange quark. The operator basis used is given below. 
	\begin{align}
		\Phi_{\mathcal{M}_{KB^*}}(x) &=&& \left[ \bar{u}(x)\gamma_i b(x)\right] \left[ \bar{d}(x)\gamma_5 s(x)\right] \nonumber \\
		& &-& \left[ \bar{u}(x)\gamma_5 s(x)\right] \left[ \bar{d}(x)\gamma_i b(x)\right], \nonumber \\
		%\end{align*}
		%\begin{align*}
		\Phi_{\mathcal{M}_{BK^*}}(x) &=&& \left[ \bar{u}(x)\gamma_5 b(x)\right] \left[ \bar{d}(x)\gamma_i s(x)\right] \nonumber \\
		& &-& \left[ \bar{u}(x)\gamma_i s(x)\right] \left[ \bar{d}(x)\gamma_5 b(x)\right], \nonumber \\
		%\end{align*}
		%
		%\begin{align*}
		\Phi_\mathcal{D}(x) &=&& \big[ 
		\left(\bar{u}(x)^T C\gamma_5 \bar{d}(x) - \bar{d}(x)^T C\gamma_5 \bar{u}(x)\right) 
		\nonumber\\
		&&&\quad \times \left(b^T(x) C\gamma_i s(x)\right)
		\big] .
		\label{eq:bs1}
	\end{align}
	In the $bs\bar u\bar d$ tetraquark sector, the relevant lowest two-particle scattering thresholds are related to $K\bar B^*$ and $\bar BK^*$ channels, which are around 350 MeV apart in the physical limit. However, in the physical limit, $K^*$ meson is unstable and can decay to $K\pi$ system and hence the lowest inelastic threshold corresponds to $\bar BK\pi$ channel, which is approximately 100 MeV above the elastic threshold corresponding to the $K\bar B^*$ channel. While we include the operators corresponding to the two-particle $\bar BK^*$ channel, we omit three-particle operators in our analysis as they go beyond the scope of our current setup and assume any effects arising out of these inelastic three-particle channels are negligible on the ground state energies we extract.

	\ul{\it $bs\bar{u}\bar{d}$ with $I(J^P)=0(0^+)$}: 
	The two operator basis that was employed in Ref. \cite{Radhakrishnan:2024ihu} in the study of isoscalar scalar $bc\bar u\bar d$ tetraquarks was utilized in this case, with the replacement of charm quark with a strange quark. Below we list the two operators used. 
	\begin{align}
		\Phi_{\mathcal{M}_{BK}}(x) &=&& \left[ \bar{u}(x)\gamma_5 b(x)\right] \left[ \bar{d}(x)\gamma_5 s(x)\right] \nonumber \\
		& &-& \left[ \bar{u}(x)\gamma_5 b(x)\right] \left[ \bar{d}(x)\gamma_5 s(x)\right], \nonumber \\
		%\end{align*}
		%
		%\begin{align*}
		\Phi_\mathcal{D}(x) &=&& \big[ \left(\bar{u}(x)^T C\gamma_5\bar{d}(x) - \bar{d}(x)^T C\gamma_5\bar{u}(x)\right) \nonumber\\
		&&&\times (b^T(x)C\gamma_5 s(x))\big].
		\label{eq:bs0}
	\end{align}
	In this channel, the lowest relevant two-particle scattering threshold is associated with the $K\bar B$ channel, scattering in the $S$-wave. The two-meson operator listed above corresponds to the lowest noninteracting level of $K\bar B$ channel, whereas the diquark-antidiquark operator is the one built from the scalar triplet color configuration of the light antidiquark. 
	
	%%%%%%%%%%%%%%%%%%%%%%%%%%%%%%%%%%%%%%%%%%%%%%%%%%%%%%%%%%%%%%%%%%%%%%%%%%%%%%%%%%%%%%%%%%%%%%%%
	%%%%%%%%%%%%%%%%%%%%%%%%%%%%%%%%%%%%%%%%%%%%%%%%%%%%%%%%%%%%%%%%%%%%%%%%%%%%%%%%%%%%%%%%%%%%%%%%
	%%%%%%%%%%%%%%%%%%%%%%%%%%%%%%%%%%%%%%%%%%%%%%%%%%%%%%%%%%%%%%%%%%%%%%%%%%%%%%%%%%%%%%%%%%%%%%%%
	
	\section{\label{sec:analysis}Analysis}
	
	The correlation matrices, $\mathcal{C}$, for all the channels studied are evaluated for the operator bases presented in Eqs. (\ref{eq:bb1}), (\ref{eq:bs1}) and (\ref{eq:bs0}) and are analyzed variationally \cite{Michael:1985ne} following the solutions of a generalized eigenvalue problem (GEVP), 
	\begin{equation}
		\mathcal{C}(t)v^{(n)}(t) = \lambda^{(n)}(t) \mathcal{C}(t_0)v^{(n)}(t).
		\label{gevp}
	\end{equation}
	Here, $\lambda^{(n)}(t)$ is the eigenvalue correlator that describes the time evolution of the $n^{th}$ lowest eigenstates with energy $E^n$. The ground state energy estimate $E^{(0)}$ can be determined from the large time behavior $\lim_{t\to\infty}\lambda^{(0)}(t) \sim A_{(0)}e^{-E^{(0)}t}$, while the overlap factors,
	\beq
	Z_i^{0}=\bra{\Omega}\Phi_i \ket{0} = \sqrt{2E^0}(V^{-1})_i^0 e^{E^{0}(t_0)/2},
	\eeq{ovp}
	can be determined from the eigenvectors matrix $V\equiv \{v^n(t)\}$, that are expected to be time independent in the large time limit, when $\mathcal{C}(t)$ is saturated by the lowest $N$ eigenstates. These overlap factors carry the information on the coupling of operators to the ground state. 
	
\begin{figure}[tb]
		\centering
		\includegraphics[width=1.0\linewidth]{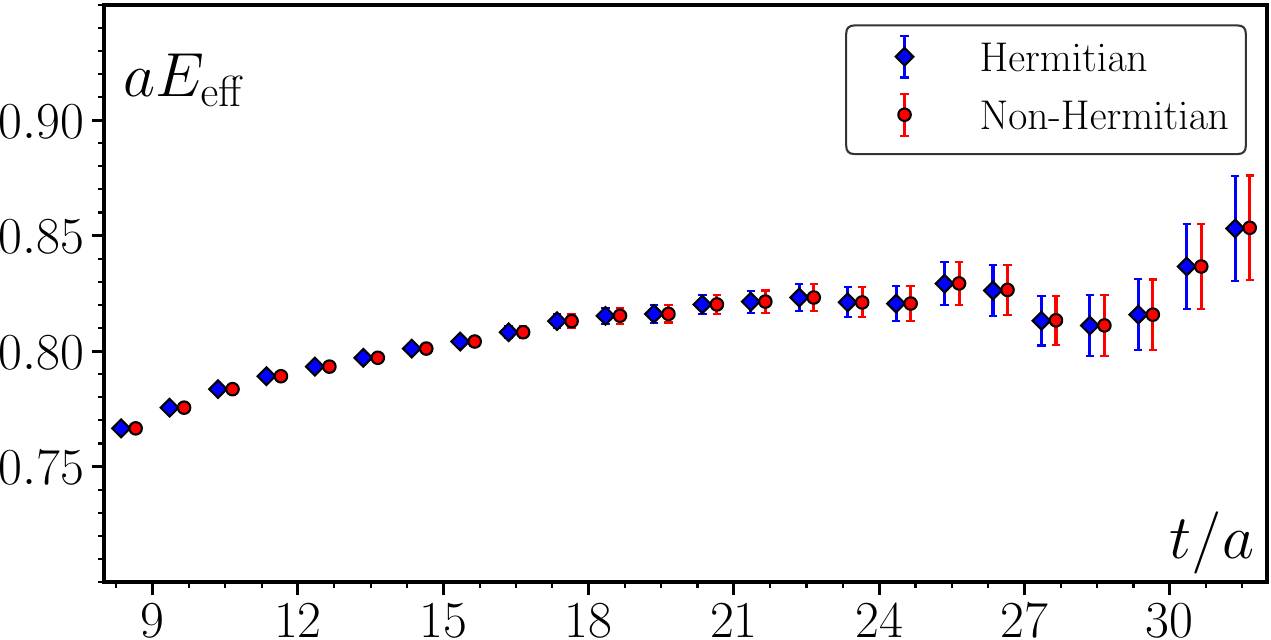}
		\caption{A comparison of the time dependence of effective masses determined from the solutions of Hermitian GEVP and Non-Hermitian GEVP applied on the asymmetric correlator setup. The results presented are for the case of the ground state in $bb\bar u\bar d$ spectrum at $M_{ps} \sim 700$ MeV on the finest ensemble. The values of $t_0$ and $t_d$ are chosen to be 6 and 19, respectively.}
		\label{HermnonHerm}
	\end{figure}

	Note that the wall-source-to-point-sink setup that we use leads to an asymmetric correlation matrix setup. With such asymmetric correlation matrices, one will need to reform the GEVP to account for the non-Hermiticity of the correlation matrices. In other words, the GEVP for these asymmetric correlation matrices can be posed in two different ways: one with a right eigenvector and the second one with a left eigenvector with the same eigenvalue \cite{Francis:2018qch}. 
	\begin{align}
		\mathcal{C}(t_{d})v_r^{(n)}(t_{d}) &= \tilde{\lambda}^{(n)}(t_{d}) \mathcal{C}(t_0)v_r^{(n)}(t_{d}) \nonumber \\
		v^{(n)\dagger}_l(t_{d})\mathcal{C}(t_{d}) &= \tilde{\lambda}^{(n)}(t_{d})v^{(n)\dagger}_l(t_{d})\mathcal{C}(t_0).
		\label{gevpnh}
	\end{align}
	Here $t_0$ is the reference timeslice for the GEVP, whereas $t_d$ is the timeslice where one does the diagonalization. After solving for the above equations at a fixed $t_0$ and $t_d$, one can determine approximate solutions at arbitrary later timeslices using the relation \cite{Francis:2018qch},
	\begin{equation}
		\tilde{\lambda}^{(n)}(t) = v^{(n)\dagger}_l(t_{d})\mathcal{C}(t)v_r^{(n)}(t_{d}).
	\end{equation}
	This assumes that the values of $t_0$ and $t_d$ are chosen such that the correlation matrix is saturated with the lowest states to a degree that the eigenvectors $v_r^{(n)}(t_d)$ and $v_l^{(n)}(t_d)$ are time-independent. We observe that the $\tilde{\lambda}^{(n)}(t)$ eigenvalues are significantly real. In other words, the ratio of the magnitudes of the imaginary part to the absolute value is less than 0.01 for all eigenvalues and at all timeslices, where the signal-to-noise ratio is good. $|\tilde{\lambda}^{(n)}(t)|$ are also observed to be consistent with the real eigenvalues $\lambda^{(n)}(t)$ of symmetrized correlation matrices (that are enforced to be Hermitian) following the standard Hermitian GEVP. This has been generally observed in all our correlator data in this work as well as in the past \cite{Junnarkar:2018twb}. In Figure \ref{HermnonHerm}, we provide an example demonstrating this equivalence in terms of effective energies defined as,
	\beq
	aE_{\mathrm{eff}} = [ln(\lambda(t)/\lambda(t+\delta t))]/\delta t .
	\eeq{meffdef}
	
\begin{figure}[tb]
		\centering
		\includegraphics[width=1.0\linewidth]{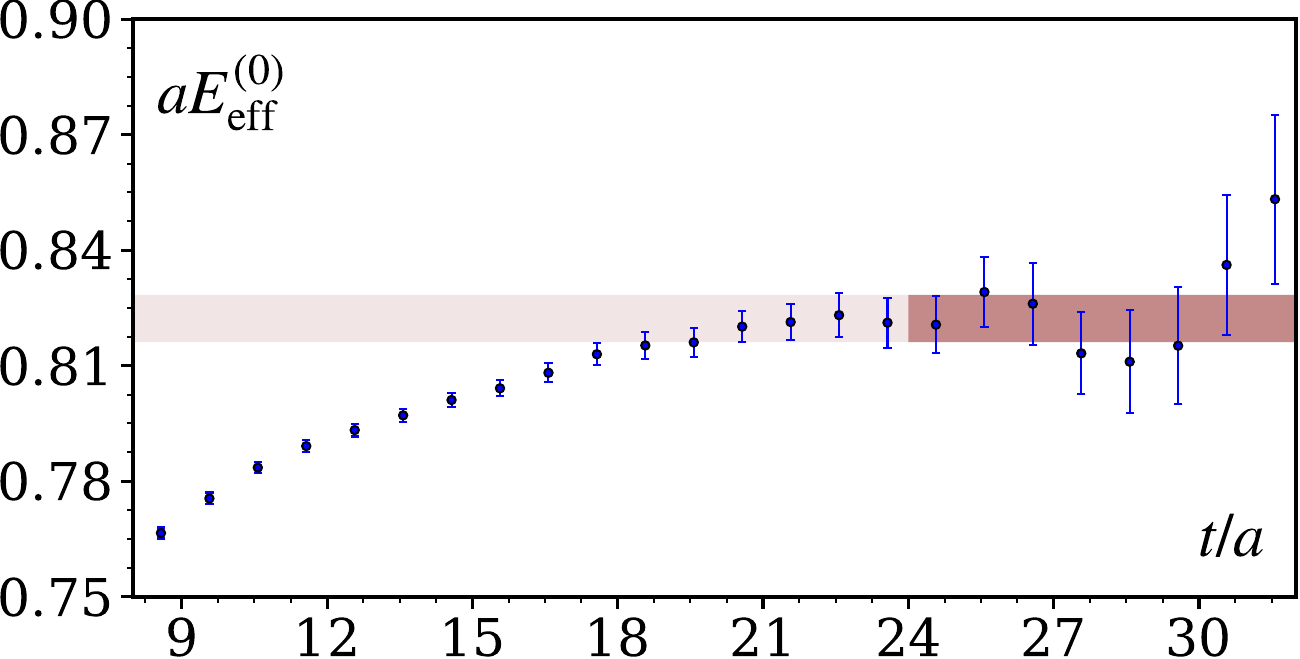}
		\caption{Effective mass plot for the ground state $\lambda^{(0)}(t)$, for $M_{ps} \sim 700$ MeV on the finest ensemble, demonstrating the signal quality. The band indicates the energy estimate obtained from fits. The presented data corresponds to the ground state in $bb\bar u\bar d$ spectrum at $M_{ps} \sim 700$ MeV on the finest ensemble.}
		\label{fg:effmass}
	\end{figure}

	The energy extraction from $\lambda^{(n)}(t)$ proceeds through an initial assessment of the ground state saturation in the $\lambda^{(n)}(t)$ from the large time plateauing of effective energy defined in \eqn{meffdef}.
	In Figure \ref{fg:effmass}, we present a demonstration of large time plateauing in $aE_{\mathrm{eff}}$ for $\lambda^{(0)}(t)$ at $M_{ps}\sim 700$ GeV on the finest lattice. Based on the time interval where plateauing is observed, the energy eigenvalues are extracted by fitting $\lambda^{(0)}(t)$ or the ratios $R^0(t)=\lambda^{(0)}(t)/\mathcal{C}_{m_1}(t) \mathcal{C}_{m_2}(t)$ with their large time exponential behavior. Here, $\mathcal{C}_{m_i}$ is the two-point correlation function for the meson $m_i$. 
	Single exponential fits to the correlator ratio $R^0(t)$ yield the estimates for energy differences $\Delta E^0 = E^0-M_{m_1}-M_{m_2}$, whereas equivalent energy differences can be computed from energy estimates from separate fits to $\lambda^{(0)}(t)$ and $\mathcal{C}_{m_i}$. The energy differences determined from lattice studies, particularly using $R^{(0)}(t)$, are empirically known to efficiently mitigate correlated noise between the interacting energy levels and the noninteracting multi-particle energies. Additionally, the additive offset inherent in energy estimates for hadrons involving bottom quarks realized with NRQCD formalism is automatically removed from the energy differences. Thus the fits to such carefully designed $R^{(0)}(t)$ directly yield energy difference estimates that are corrected for this NRQCD offset. 
	
	\begin{figure}[tb]
		\centering
		\includegraphics[width=1.0\linewidth]{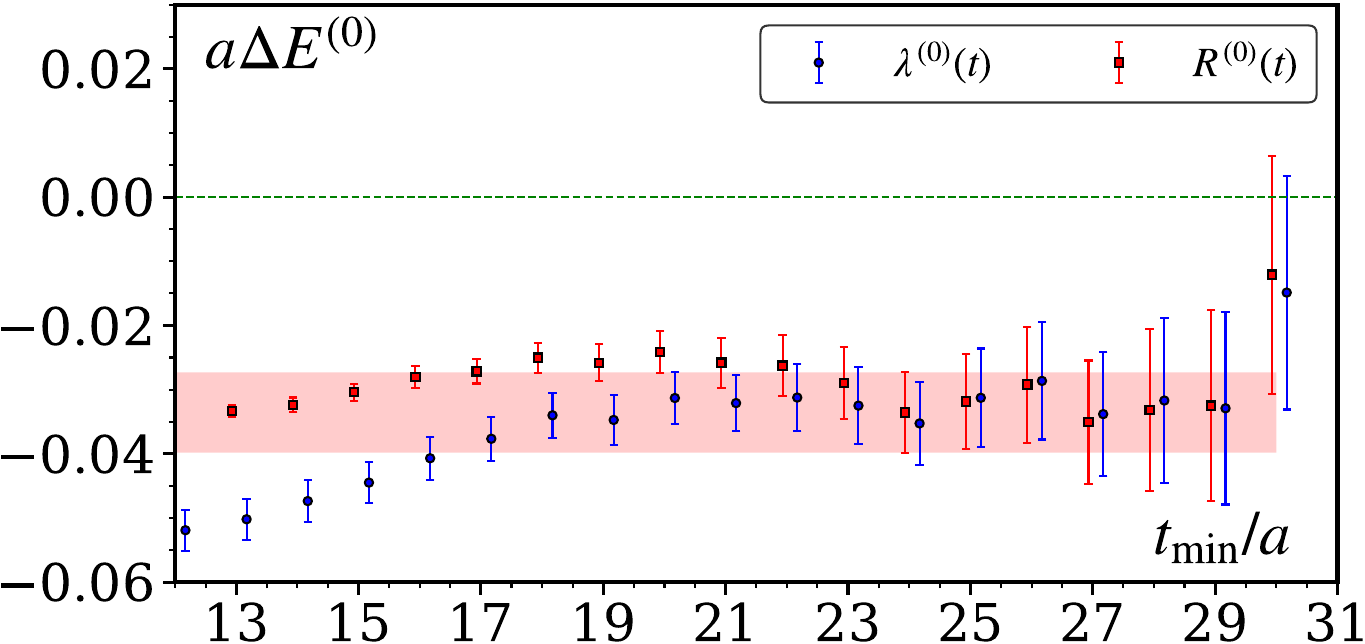}
		\caption{$t_{\text{min}}$ dependence of the ground state energy estimates, presented in terms of the energy splittings $a\Delta E^{(0)}$, obtained from the fits to $\lambda^{(0)}(t)$ and $R^{(0)}(t)$ for the ground state in $bb\bar u\bar d$ spectrum at $M_{ps} \sim 700$ MeV on the finest ensemble. The red band represents 1$\sigma$ uncertainty estimated via a bootstrap resampling procedure.}
		\label{fg:tmin}
	\end{figure}
	In general, we make a comparative study of the fits (to $\lambda^{(0)}(t)$ and $R^{(0)}(t)$) to make sure that our final choice of fitting time intervals is not affected by any conspired cancellation of signals in the correlators leading to misidentification of a fake plateau as the real ground state energy. Such a consistency check between the estimates from $\lambda^{(0)}(t)$ and $R^{(0)}(t)$ would be helpful in identifying the true ground state plateau. In Figure \ref{fg:tmin}, we present such a comparison of energy differences determined from fits to $\lambda^{(0)}(t)$ and $R^{(0)}(t)$. It is transparent from the figure that at large times, $\Delta E^0$ determined from either procedure leads to consistent estimates within statistical uncertainties. We observe such consistency in other correlator data as well. Our results for the finite volume energy eigenvalues are based on the fits to $R^{0}(t)$.

	The choice of fitting time intervals is arrived at based on a study of $t_{min}$ (the early time boundary of the fitting time interval) dependence of the fit estimates for a fixed $t_{max}$ (the late time boundary of the fitting time interval) chosen considering the signal quality degradation in $\lambda^{0}(t)$. Once again Figure \ref{fg:tmin} shows such $t_{min}$ dependence of the energy as well as energy differences extracted that guides in arriving at a conservative choice of fitting time interval, where the two different procedures are found to agree. 
	
	Considering the asymmetric nature of our correlation functions, we make further checks on our $t_{min}$ choices following alternative smearing procedures for the quark fields at the sink. In the asymmetric scenario, one expects deviations in the effective mass from a conventional falling-from-above feature as a result of strictly positive contributions from higher excitations in a spectral decomposition, modulo the statistical noise. Towards the symmetric limit, one expects the restoration of the conventional expectation on the approach to large time behavior. While approach to the large-time behavior could be different between the symmetric and the asymmetric setups, the energy estimates in the large time limit, where the contributions from higher excitations have died out, should be consistent between different setups. 
	
\begin{figure}[tb]
		\centering
		\includegraphics[width=\linewidth]{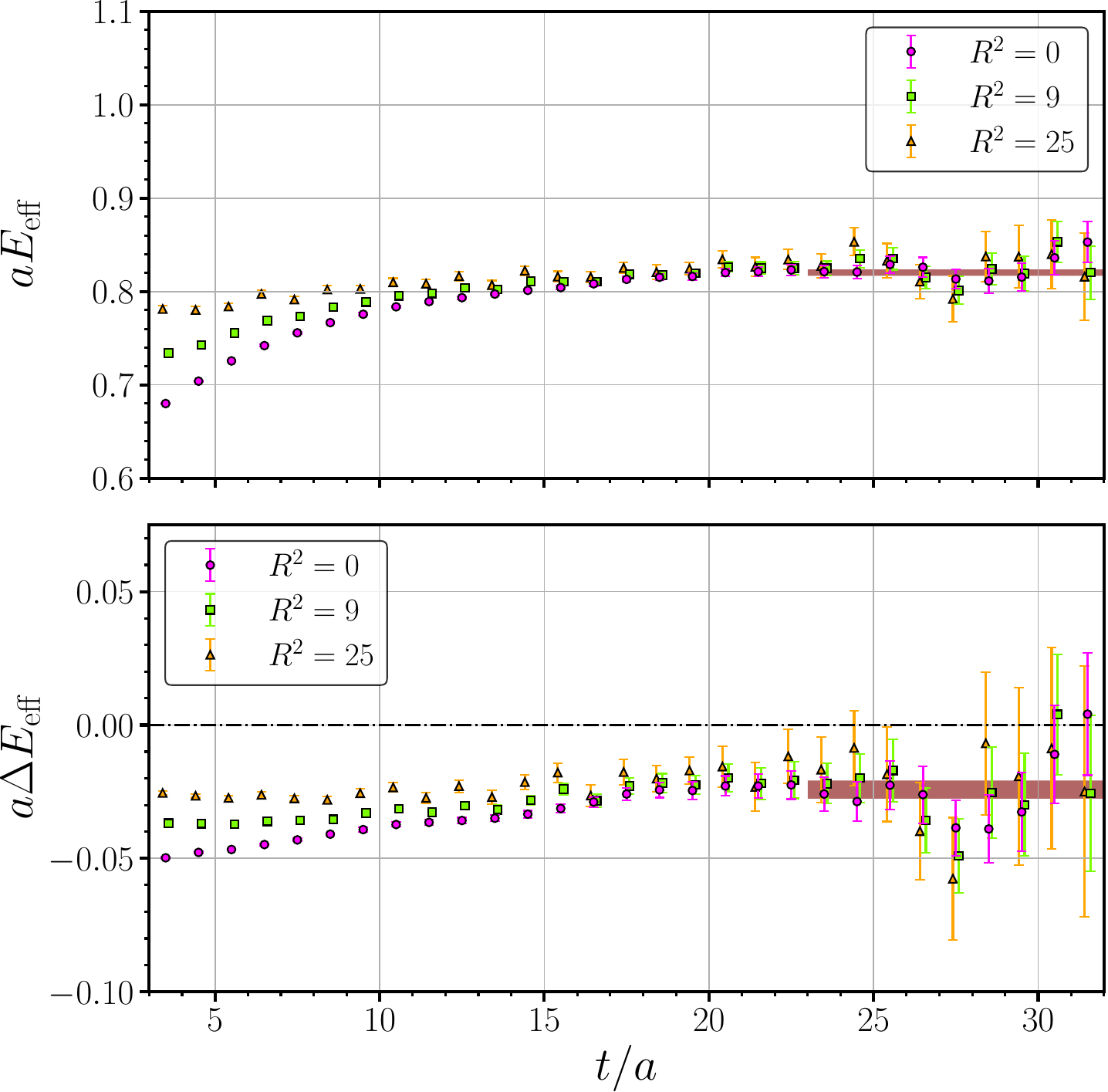}
		\caption{Comparison of the effective energy (top) and effective energy splitting (bottom) for the ground state, determined using three different smearing radii applied to the quark fields at the sink timeslice. The legend denotes the squared smearing radius in units of the lattice spacing \cite{Hudspith:2020tdf}. The maroon horizontal band represents the final fit estimates for the energy and energy splitting. These results correspond to the ground state in the $bb\bar u\bar d$ spectrum at $M_{ps} \sim 700$ MeV on the finest ensemble.}
		\label{boxsnk_plot}
	\end{figure}

	We investigate the approach to and the energy estimated at the large time limit across different sink smearing radii, to build confidence in our final results. In Figure \ref{boxsnk_plot}, we present a comparison of the effective mass corresponding to the solutions of asymmetric correlation matrices using wall-source point-sink setup with those determined from correlation matrices using a wall-source box-sink setup \cite{Hudspith:2020tdf} with varying box radii, which asymptotically approaches the symmetric limit. The results are presented for the $bb\bar u\bar d$ case at $M_{ps}\sim700$ MeV on the finest ensemble. The deviations from the conventional expectations are evident in the asymmetric setups and gradually disappear approaching the symmetric limit. This is consistent with observations in our previous calculations \cite{Junnarkar:2018twb, Mathur:2022ovu, Padmanath:2023rdu, Radhakrishnan:2024ihu} as well as in Ref. \cite{Hudspith:2020tdf}. The energy and energy difference estimates at the large time limit can also be observed to be consistent between different setups adding credibility to our final choice for fitting time intervals. In the $bb\bar u\bar d$ case, we observe that all these setups consistently indicate negative energy shifts suggesting attractive interactions between the $B$ and $B^*$ mesons. 
	
	%%%%%%%%%%%%%%%%%%%%%%%%%%%%%%%%%%%%%%%%%%%%%%%%%%%%%%%%%%%%%%%%%%%%%%%%%%%%%%%%%%%%%%%%%%%%%%%%
	%%%%%%%%%%%%%%%%%%%%%%%%%%%%%%%%%%%%%%%%%%%%%%%%%%%%%%%%%%%%%%%%%%%%%%%%%%%%%%%%%%%%%%%%%%%%%%%%
	%%%%%%%%%%%%%%%%%%%%%%%%%%%%%%%%%%%%%%%%%%%%%%%%%%%%%%%%%%%%%%%%%%%%%%%%%%%%%%%%%%%%%%%%%%%%%%%%
	\section{\label{sec:finite_volume}Finite volume energy eigenvalues}
	
\begin{figure}[htbp]
		\centering
		\includegraphics[width=1.0\linewidth]{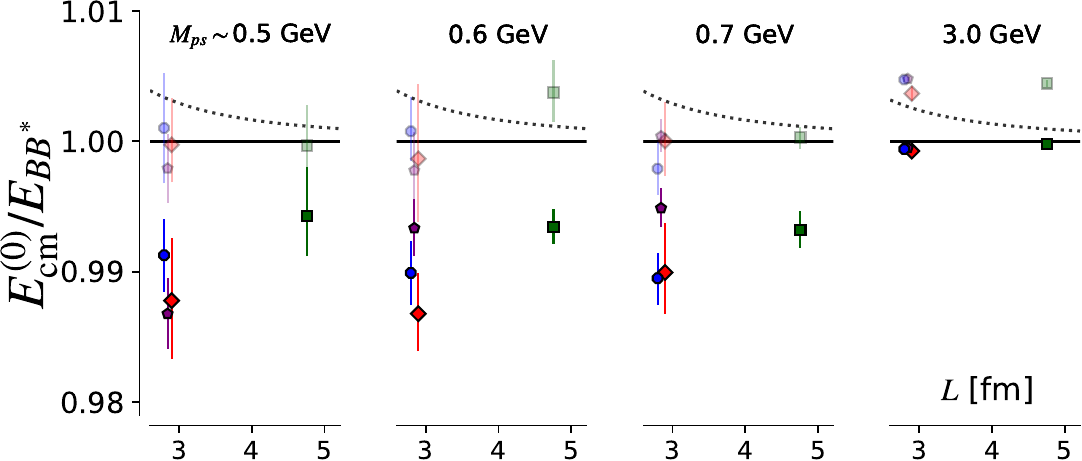}
		\caption{The ground state and first excited state energies of the $0(1^+)$ $bb\bar{u}\bar{d}$ channel in finite volume, shown across different pseudoscalar masses $(M_{ps})$ in separate vertical panels. The y-axis represents the center-of-mass energies, normalized to the nearest two-body elastic threshold $BB^*$. The first excited noninteracting two-meson level with a single unit of lattice momenta for the mesons involved is shown in the dotted curve. The x-axis denotes the spatial extent of each ensemble.}
		\label{nrqcd_bb}
	\end{figure}

\begin{figure}[htbp]
		\centering
		\includegraphics[width=1.0\linewidth]{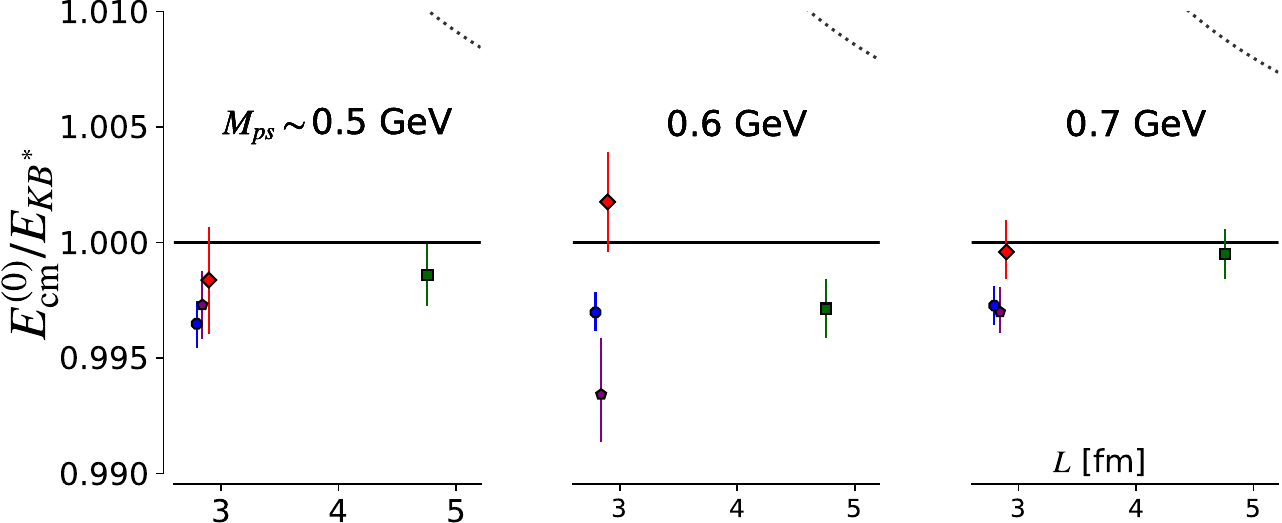}
		\caption{Same as in Figure \ref{nrqcd_bb}, but for the ground state energies in the $0(1^+)$ $bs\bar{u}\bar{d}$ channel. The energy levels are presented in units of the energy of $KB^*$ threshold.}
		\label{nrqcd_bsav}
	\end{figure}

\begin{figure}[htbp]
		\centering
		\includegraphics[width=1.0\linewidth]{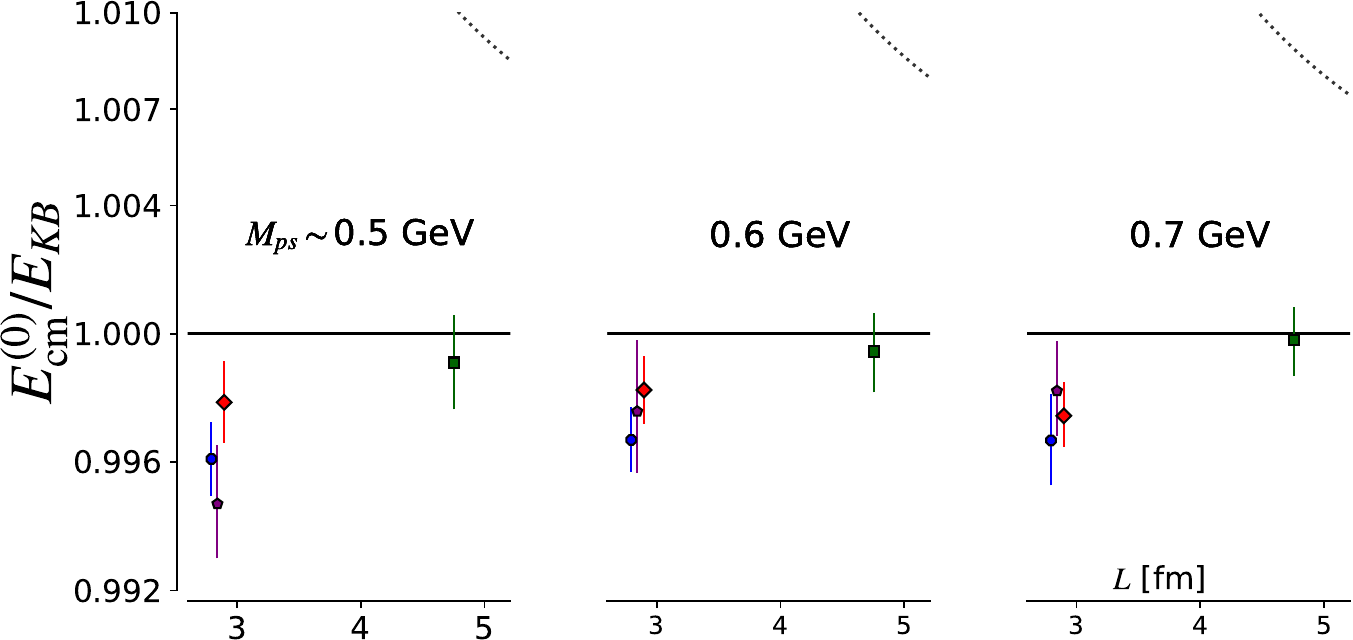}
		\caption{Same as in Figure \ref{nrqcd_bb}, but for the ground state energies in the $0(0^+)$ $bs\bar{u}\bar{d}$ channel. The energy levels are presented in units of the energy of $KB$ threshold.}
		\label{nrqcd_bssc}
	\end{figure}

	In Figures \ref{nrqcd_bb}, \ref{nrqcd_bsav}, and \ref{nrqcd_bssc}, we present the ground state eigenenergies in the finite volume for the axialvector $bb\bar u\bar d$, axialvector $bs\bar u\bar d$ and scalar $bs\bar u\bar d$ systems, respectively. For the axialvector $bb\bar u\bar d$, we additionally present the first excited eigenenergies using markers with transparency. The eigenenergies are presented in units of the respective elastic threshold energies $E_{m_1m_2}$ in each case studied, whereas the $x$-axis represents the spatial lattice extension of the ensembles used. In the rest of this section, we discuss the details of these figures and the observations we make from them, based on which we proceed to the scattering analysis, wherever possible. 
	
	Note that the energy estimates from fits to the eigenvalue correlators are not free of the additive offsets related to the NRQCD dynamics of the bottom quarks. However, these offsets are accounted for in our final results, presented in \ref{nrqcd_bb}, \ref{nrqcd_bsav}, and \ref{nrqcd_bssc}, determined based on the eigenenergy splittings $\Delta E$ extracted from the ratio correlators $R^0(t)$, defined in the previous section. To this end, we consider the ratios $R^0(t)$ built with respect to the noninteracting correlator corresponding to the elastic two meson scattering channel in each case studied. For the $bb\bar u\bar d$ case, this is $BB^*$, whereas for the $bs\bar u\bar d$ axialvector and scalar cases, the elastic channels are $KB^*$ and $KB$, respectively. The eigenenergies are the reconstructed from $\Delta E$ as $E = \Delta E + M_{m_1} + M_{m_2}=\Delta E +E_{m_1m_2}$, where $M_{m_i}$ is the corresponding lattice estimate for the mass of meson $m_i$ within the same setup. Thus in the  $bb\bar u\bar d$ case, $E_{m_1m_2}=E_{BB^*}$, whereas for the $bs\bar u\bar d$ axialvector and scalar cases, $E_{m_1m_2}=E_{KB^*}$ and $E_{KB}$, respectively. Note that the meson masses, involved here, are also subject to the additive offset related to the NRQCD bottom quark dynamics. Hence, the NRQCD-offset corrected lattice estimates for mass of the relevant bottom mesons are evaluated as  $\tilde{M}_{B^{(*)}} = M_{B^{(*)}}$ $- 0.5\overline M^{\bar bb}_{lat} + 0.5 \overline M^{\bar bb}_{phys}$, where $\overline M^{\bar bb}_{lat}(\overline M^{\bar bb}_{phys})$ is the spin averaged mass of the $1S$ bottomonium on the lattice (experiments). The factor 0.5 in front of $\overline M^{\bar bb}_{\ast}$ above accounts for the NRQCD offset arising out of a single valence bottom quark in the bottom mesons. 
	
	In Figure \ref{nrqcd_bb}, negative energy shifts in the ground states with respect to the elastic $BB^*$ threshold are evident across all light quark mass cases studied, suggesting attractive interactions between the $B$ and $B^*$ mesons. The first excited states extracted from the simulation are also presented, and they are generally consistent with the elastic threshold. We also indicate the lowest omitted noninteracting level, which involves one lattice unit of relative momenta between the mesons giving zero total momentum of the system. Another observable effect is the gradual decrease in the binding energy in units of the elastic threshold energy with increasing pseudoscalar meson mass $M_{ps}$, suggesting decreasing strength of the attractive interactions. A nontrivial dependence on the lattice spacing ($a$) can also be observed on the ground state energies on the small volume ensembles, which we account for in the analysis discussed below. 
	
	In Fig. \ref{nrqcd_bsav}, we present the finite volume ground state energy estimates in units of the elastic threshold $E_{KB^*}$ in the axialvector $bs\bar u\bar d$ channel. No strong variation can be observed in the energy shifts with respect to the threshold as a function of $M_{ps}$. The green square corresponding to the large volume ensemble clearly suggests consistent energy estimates with the threshold across all $M_{ps}$, suggesting negligible interactions if any. This is consistent with phenomenological expectation for the doubly heavy system, which suggests lighter the heavy diquark reduced mass, the weaker the binding in the doubly heavy tetraquark system \cite{Francis:2016hui,Czarnecki:2017vco}. This observation also goes in line with the pattern of binding energies observed in our previous investigations: $\mathcal{O}(100 $MeV$)$ binding in the $bb\bar u\bar d$ system, and $\sim40$ MeV in $bc\bar u\bar d$ system. Given the consistency with the threshold energy of the finite volume ground states, we have performed a simple minded amplitude analysis allowed by the degrees of freedom available.  
	
	Note that the operators used are local two-meson interpolators and diquark-antidiquark operators, which need not faithfully reproduce the true elastic excitations as pointed out in Ref. \cite{Padmanath:2023rdu}. We note that the finite volume ground states in our setup are primarily determined by the two-meson operators of type $KB^*$, whereas the first excitations are dominantly controlled by the $BK^*$-type two-meson interpolators. The lattice-extracted first excitation energies are close to the inelastic $BK^*$ threshold that is very high in energy, and hence they clearly do not represent the true elastic excitations in the $KB^*$ channel. Note that the use of wall-source setup and additional box-sink setup ensures the correct identification of the finite volume ground state energies as demonstrated in the previous section and our previous investigations. Hence we limit ourselves to the ground state energies in the rest of the discussion. 
	
	Negative energy shifts with varying $M_{ps}$ could be argued in the scalar $bs\bar u\bar d$ channel presented in Fig. \ref{nrqcd_bssc}, although statistically insignificant and consistent with zero. The ground state energies are presented in units of the elastic threshold energy $E_{KB}$. Similar to the axialvector case, the energy shift in the large volume ensemble remains consistent with the threshold energy across all $M_{ps}$, indicating only remnant weak interactions, if exist. The lowest inelastic two-meson threshold corresponds to $K^*B^*$, which is higher in energy than the lowest inelastic threshold corresponding to the four meson system $KB\pi\pi$ that is $\sim$280 MeV above $E_{KB}$ at the physical point. In this study, we have $M_{ps}$ down to only 500 MeV, such either of these inelastic thresholds is sufficiently higher up to consider the scalar channel within an elastic approximation in further analysis that we perform. Note that in either of the $bs\bar u\bar d$ cases, we ignore the case $M_{ps}=3.0$ GeV. 
	
	We iterate that, in all these cases we limit the use of only the finite volume ground state energies for scattering analysis, as the excited levels extracted from the correlation functions may not be associated with the lowest elastic excitations. In our previous publications, the ground state energies in the rest frame are empirically observed to be reproduced reliably with the wall-smearing setup. As demonstrated in the previous sections, we employ various cross checks that aid us mitigate the excited state contamination and reliably identifying the true ground state plateau. 
	
	It was pointed out by the authors of Ref. \cite{Du:2023hlu}, that the studies of doubly heavy tetraquark systems could be influenced by left hand nonanalyticities arising from long range interactions due to meson exchange processes in the crossed channels. Similar to the case of $DD^*$ scattering addressed in Ref. \cite{Du:2023hlu}, the nearest left hand branch point to the $BB^*$ threshold corresponds to a pion exchange process in the $u$-channel. In the physical world, following the same arguments and formulae (Eq. (8) in Ref. \cite{Du:2023hlu}), the momentum-squared at the left-hand cut branch point turns out to be $\mathcal{O}(10^{-4}E_{BB^*})$, which is just $\sim1$ MeV below the $BB^*$ threshold. At the lightest pseudoscalar meson mass we work with (500 MeV), this is $\sim12$ MeV below $E_{BB^*}$. Considering the binding energy estimates for $T_{bb}$ ($\mathcal{O}(100$ MeV$)$) predicted by the phenomenological investigations as well as those based on lattice QCD simulations, all of which ignore any related effects, this is a matter of concern for future studies of $T_{bb}$. Like in the previous investigations, we continue to ignore the effects of pion exchange left-hand cut effects, as this goes beyond the scope of this work.

	In the context of the axialvector $bs\bar u\bar d$ channel, an offshell light pseudoscalar meson exchange will introduce coupling between the $KB^*$ and $BK^*$ channels, which is another technicality that is beyond the scope of our understanding. An exchange process limiting to a single channel ($KB^*$) can happen only if the exchanged meson is a pseudoscalar bottom-strange $B_s$ meson. Such an exchange is not long ranged considering the mass of the $B_s$ meson, and the corresponding left-hand cut branch point would feature significantly outside the view of the spectrum plot provided in Figure \ref{nrqcd_bsav}. A similar situation arises also in the case of $KB$ scattering, where the exchange particle has to be a $B_s^*$ meson, which would lead to a left-cut branch point far below the $y$-axis range presented in the Figure \ref{nrqcd_bssc}. Note that similar branch points associated with other allowed multi-particle exchanges (two-pion exchanges) in the crossed channels may feature much closer to the two-particle threshold, than the above referred single-particle cross-channel exchange process. Theoretical studies on the effects of such two-particle cross-channel exchanges that lead to left-hand nonanalyticities are limited \cite{Chacko:2024ypl}.

	\section{\label{sec:ampana}Scattering amplitudes}
	
	Following the extraction of ground state energies, we extract the $S$-wave amplitudes in the channels studied within elastic approximation following the finite-volume two-particle spectrum quantization prescription \ala~L\"uscher and its generalizations \cite{Luscher:1990ux, Briceno:2014oea}. For all the three cases we investigate, within an elastic scattering assumption the $S$-wave phase shift $\delta_{l=0}(k)$ can be extracted from finite-volume spectral energies via the quantization relation $p\cot[\delta_0(p)] = 2Z_{00}[1;(\frac{pL}{2\pi})^2]/(L\sqrt{\pi})$, where $p$ is the momentum of the scattering particles in their center of momentum frame. $p$ is related to the total energy $E_{cm}$ through $4sp^2 = (s-(M_{1}+M_{2})^2)(s-(M_{1}-M_{2})^2)$, where $s=E_{cm}^2$ is the Mandelstam variable and $M_{i}$ is the mass of the scattering particle. We enforce the quantization constraint to extract the energy dependence of the amplitude following the procedure discussed in Appendix B of Ref. \cite{Padmanath:2022cvl}. Following the extraction of the amplitude, we search for possible near threshold poles representing any bound states in the amplitude. Given the constraints on going beyond a constant parametrization in $p^2$, we would not be able to access any information on resonances that might exist. A real bound state exists in the $S$-wave amplitude, when $p{\mathrm{cot}}\delta_0 = +\sqrt{-p^2}$. 
	
	\begin{table}[h]
		\setlength{\tabcolsep}{0.9mm}{
			\centering
			\renewcommand\arraystretch{1.9} 
			\begin{tabular}{|c|c|c|c|c|}
				\hline
				$M_{ps}$ [GeV] & $\chi^2$/d.o.f & $A^{[0]}/E_{BB^*}$ &  $A^{[1]}/E_{BB^*}$ &$a_0$ [fm] \\
				\hline
				0.5 & 3.267/2 & $-0.074(^{+19}_{-20})$ & $0.03(^{+22}_{-20})$ &$ 0.25(^{+9}_{-5})$ \\
				0.6 & 4.65/2 & $-0.072(^{+18}_{-17})$ & $0.09(17)$ &$0.26(^{+8}_{-6})$ \\
				0.7 & 4.67/2 & $-0.076(17)$ & $0.17(1)$ & $0.24(^{+8}_{-4})$\\
				3.0 & 0.77/2 & $-0.012(^{+5}_{-6})$ & $-0.02(^{+8}_{-7})$ &$1.44(^{+1.04}_{-0.47})$\\
				\hline
		\end{tabular}}
		\caption{The best fit parameter values for $BB^*$ amplitude based on the parametrization in \eqn{zerorange} for different pseudoscalar masses $M_{ps}$ listed in the first column. The second column indicates the quality of fits in terms of the chi-square per degrees of freedom. The third and fourth columns contain the parameter values presented in units of the energy of $BB^*$ threshold. The fifth column gives the continuum extrapolated scattering length $a_0$ in physical units, determined from the best fit parameter value $A^{[0]}$.}
		\label{tab:bbudfit}
	\end{table}
	
	\begin{figure}[htbp]
		\centering
		\includegraphics[width=\linewidth,height=11cm]{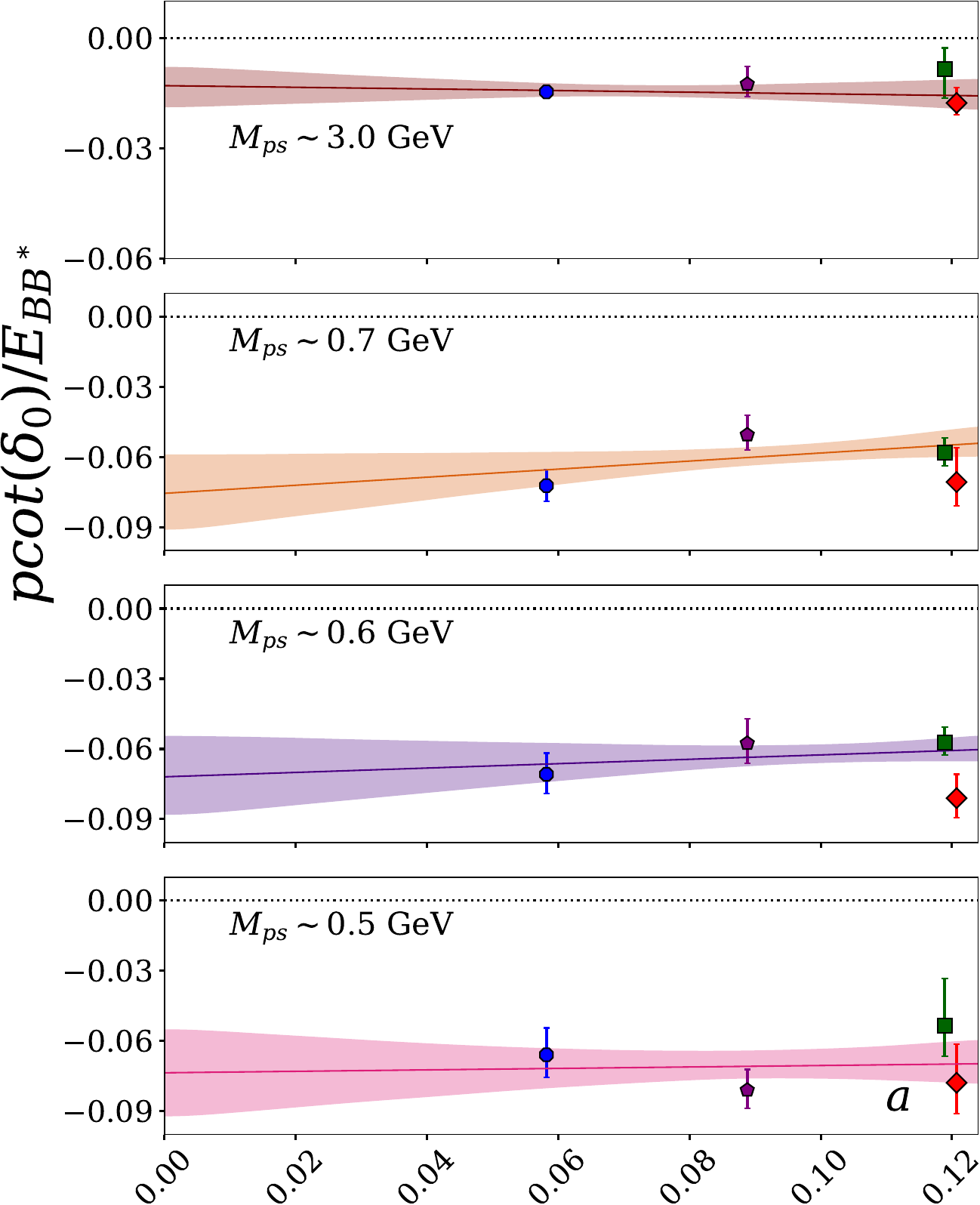}
		\caption{$p \cot \delta_0$ normalized to the elastic threshold energy $E_{BB^*}$ as a function of lattice spacing $a$ for various pseudoscalar meson masses ($M_{ps}$). The marker-color convention is listed in Table \ref{ensemb_table}, and represents the interacting finite-volume data. The colored band illustrates the continuum extrapolation fit results, with 1$\sigma$ uncertainties.}
		\label{bba}
	\end{figure}
	
	\subsection{$bb\bar u\bar d$ tetraquarks}
	\begin{figure}[htb]
		\centering
		\includegraphics[width=\linewidth,height=10cm]{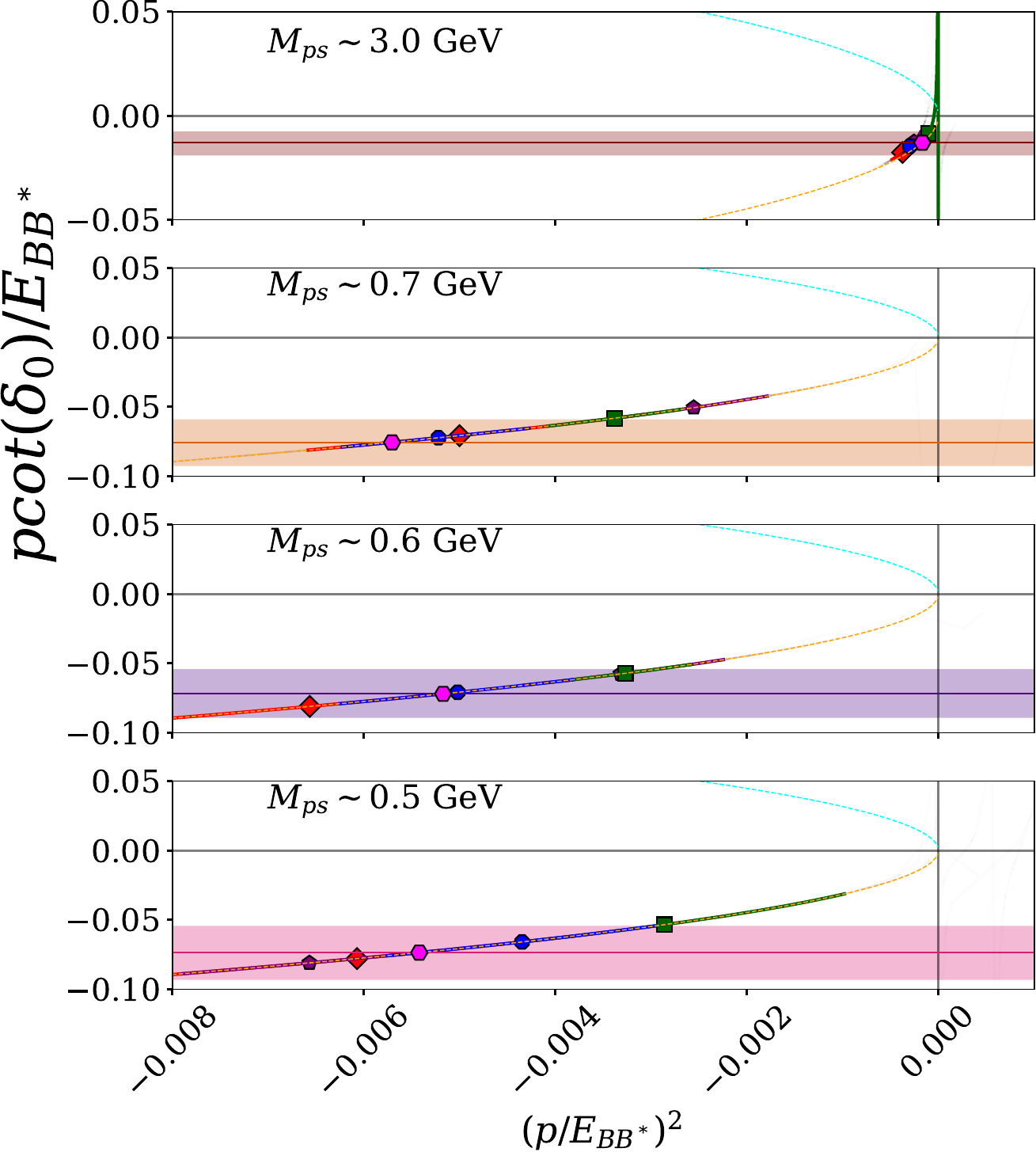}
		\caption{$p \cot \delta_0$ versus $p^2$ in the units of the elastic threshold $E_{BB^{*}}$ for all the $M_{ps}$ studied. The horizontal band represents the negative of continuum extrapolated inverse scattering length. The magenta symbols refer to the subthreshold pole in the extracted continuum extrapolated scattering amplitude, represented by the bands at respective $M_{ps}$ values. The unitary parabola shown in cyan and orange curves represents the constraint curves for the existence of subthreshold poles along the real energy axis. }
		\label{bbere}
	\end{figure}
\begin{figure}[htbp]
		\centering
		\includegraphics[width=\linewidth,height=3.5cm]{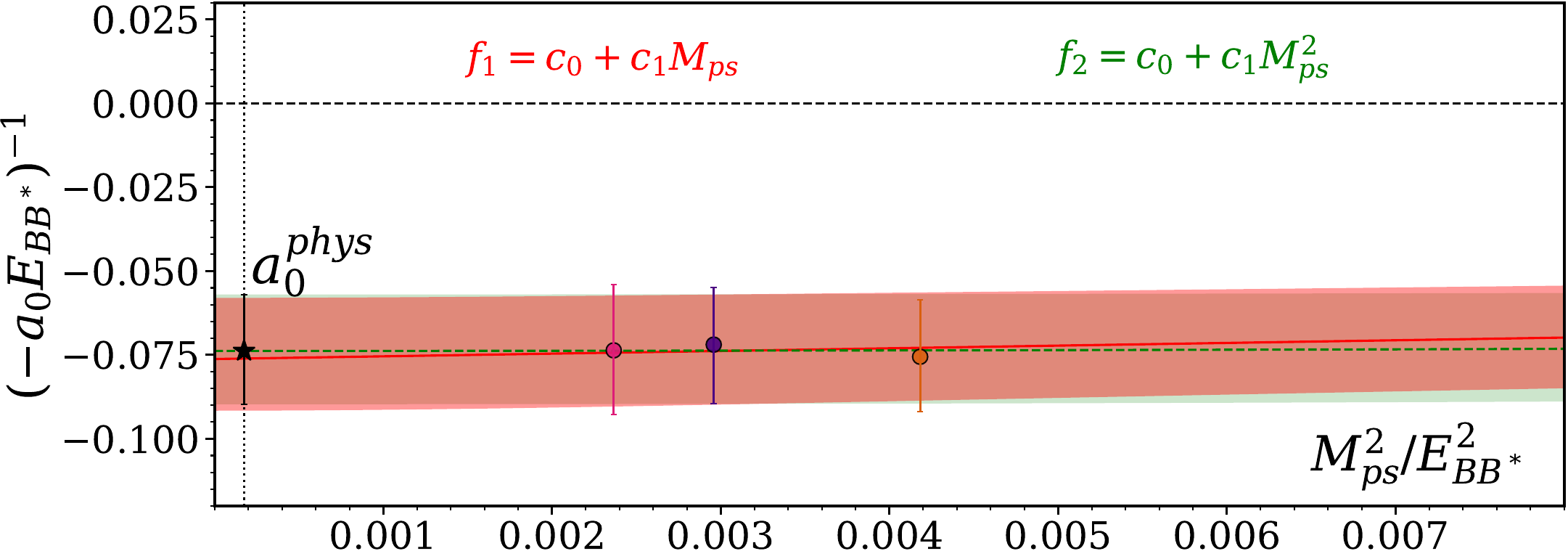}
		\caption{Continuum extrapolated $p \cot \delta_0$ estimates of the $BB^{*}$ system as a function of $M^2_{ps}$ in the units of $E_{BB^{*}}$. The vertical dotted line close to the y-axis spine represents the physical pion mass $M^{phys}_{\pi}$. The star symbol at the physical pion mass represents the amplitude at the physical point $(a_0^{phys}E_{BB^*})^{-1}$.}
		\label{chiral_bb}
	\end{figure}

	In Table \ref{tab:bbudfit}, we present the L\"uscher-based amplitude fit results from the finite-volume $BB^*$ scattering data, where we assumed a zero range approximation for the amplitude and a linear dependence on the lattice spacing $a$ to account for the cut off effects, like in our previous publications \cite{Padmanath:2023rdu, Radhakrishnan:2024ihu}. With this assumption, 
	\beq
	p~\cot(\delta_0) = A^{[0]}+a \cdot A^{[1]},
	\eeq{zerorange}
	where $A^{[0]}=-1/a_0$ is the negative of inverse scattering length in the continuum limit. The lattice spacing dependence of the extracted $p~\cot(\delta_0)$ in units of $E_{BB^*}$ is presented in Figure \ref{bba}, whereas the continuum estimate is presented along with the respective lattice degrees of freedom in the standard $p~\cot(\delta_0)$ versus $p^2$ format in Figure \ref{bbere}. The crossing of the continuum extrapolated amplitude with the (real/virtual) bound state pole constraint curves given by the (orange/cyan) parabolic curve indicates the position of any below threshold pole. These subthreshold pole positions for each $M_{ps}$ are represented by the magenta hexagons. In Figure~\ref{chiral_bb}, we present the continuum extrapolated amplitudes as a function of the $M_{ps}$ involved. Note that the amplitude is stable with respect to the $M_{ps}$ across the lowest three values, indicating hardly any dependence on the light meson mass involved in the $BB^*$ interactions. The invariance of the amplitude on the $M_{ps}$ involved suggests negligible effects on the chiral extrapolated amplitude with different fit forms, such as a constant or a quadratic in $M_{ps}$ inspired from chiral effects or a linear form inspired from Heavy Quark Effective theory, compared to the respective statistical uncertainties. The resultant chiral extrapolated amplitude suggests a bound state pole in the physical $BB^*$ amplitude with a binding energy of $\Delta E_{T_{bb}}(1^+)=-116(^{+30}_{-36})$ MeV. One may argue the largeness of $M_{ps}$ values considered in such chiral extrapolations. The variation in this estimate if one had chosen the amplitude at the lightest $M_{ps}$ we investigate or with a na\"ive continuum extrapolation of the binding energy estimates at the lowest $M_{ps}$, is well within a few MeV about 116 MeV with reasonable reduced chi-square values between 1 and 1.5. Note that the observed independence of the scattering length with $M_{ps}$ might be a consequence of the inability to resolve the valence quark mass dependence with the available statistics, and towards the chiral limit there could be systematic uncertainities that one might be blind about. In the future, we hope to extend our investigations to lighter $M_{ps}$ values to further quantify this. We also observe that the amplitude and the resultant binding energy estimate at the physical point are hardly influenced by the inclusion of the finite-volume data at $M_{ps}\sim3.0$ GeV. 
	
	Note that the above extraction of amplitude from the finite-volume results is based on L\"uscher-based quantization conditions for $2\rightarrow2$ scattering. As pointed out in the previous section, this procedure remains invalid close to and below the left-hand branch point arising out of single-pseudoscalar meson exchange in the cross-channel process, which in this case happens to be $\sim12$ MeV below the $BB^*$ threshold. This logarithmic cut running from this left-hand branch point to $-\infty$ leads to complexified amplitude below the \lhc~branch point \cite{Du:2023hlu, Collins:2024sfi}. This implies that the existence of bound states with binding energy $\mathcal{O}(100$ GeV$)$ in this situation can happen only if the imaginary part of $p\cot(\delta_0)$ goes to zero, while the corresponding real part crosses the bound state pole constraint $-i\sqrt{-p^2}$ at any given energy. Given this observation, the claims on a bound $T_{bb}$, ignoring any effects of this \lhc,~ have to be taken with a grain of salt. Any quantified comments based on the finite-volume data we have on how such fine tuning can happen and how that would lead to a real bound state is beyond the scope of this work and is a subject for future studies. 
	
	\subsection{$bs\bar u\bar d$ tetraquarks}
	\begin{table}[h]
		\setlength{\tabcolsep}{0.9mm}{
			\centering
			\renewcommand\arraystretch{1.9} 
			\begin{tabular}{|c|c|c|c|}
				\hline
				$M_{ps}$ [GeV] & $\chi^2$/d.o.f & $A^{[0]}/E_{KB^*}$ &  $A^{[1]}/E_{KB^*}$ \\
				\hline
				0.5 & 0.3/2 & $-0.02(3)$ & $0.24(^{+51}_{-41})$ \\	
				0.6 & 8.59/2 & $-0.002(^{+16}_{-26})$ & $-0.07(^{+46}_{-19})$ \\
				0.7 & 1.62/2 & $-0.029(^{+26}_{-40})$ & $0.40(^{+67}_{-31})$ \\
				\hline
		\end{tabular}}
		\caption{Same as in Table \ref{tab:bbudfit}, but for the axialvector $bs\bar u\bar d$ channel. The best fit parameter values presented in the third and fourth columns are in units of the energy of $KB^*$ threshold. Since the inverse scattering lengths are consistent with zero, we refrain from quoting any estimates for the scattering lengths in physical units. }
		\label{tab:bsudavfit}
	\end{table}
	\begin{figure}[htbp!]
		\centering
		\includegraphics[width=\linewidth,height=8cm]{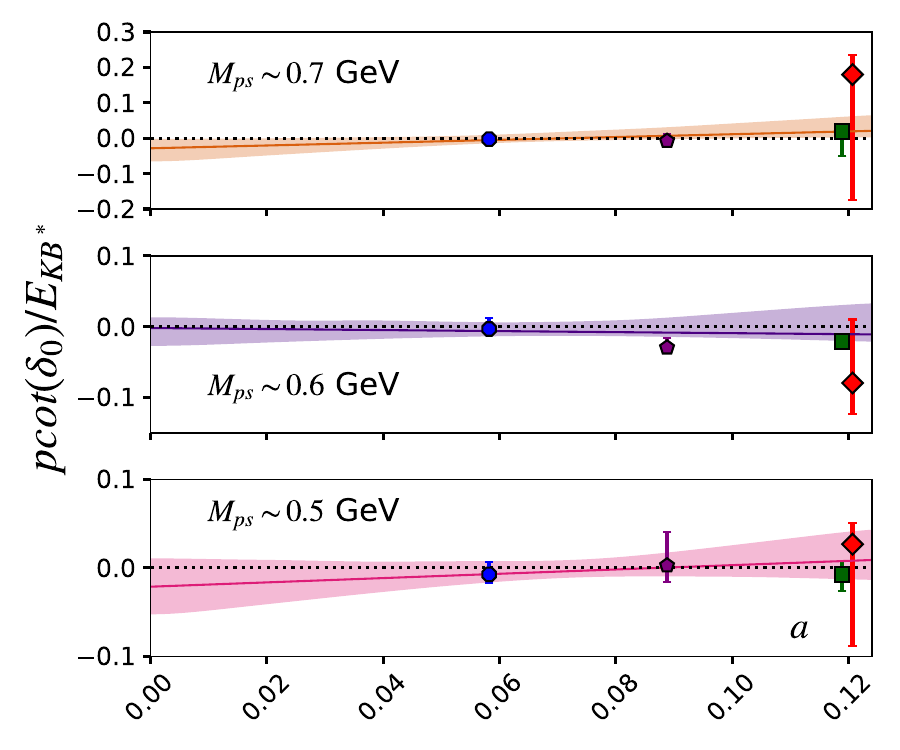}
		\caption{Same as in Figure \ref{bba}, but for the axialvector channel in $KB^*$ scattering. $pcot(\delta_0)$ is presented in units of the energy of $KB^*$ threshold.}
		\label{bsava}
	\end{figure}

\begin{figure}[htbp!]
		\centering
		\includegraphics[width=\linewidth]{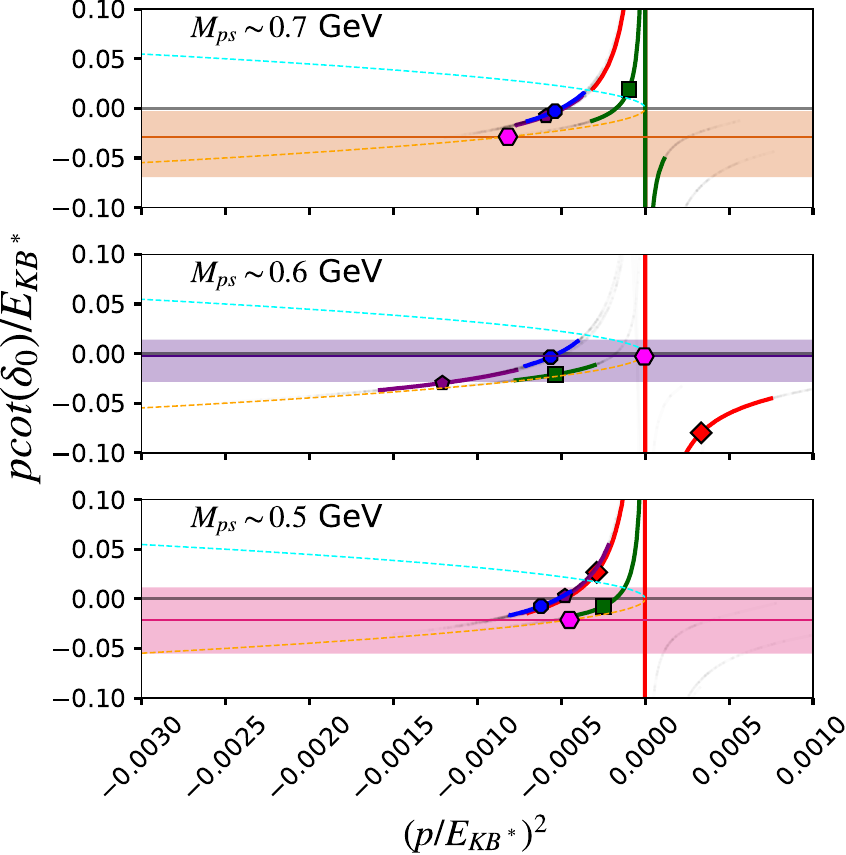}
		\caption{Same as in Figure \ref{bbere}, but for the axialvector channel in $KB^*$ scattering. $pcot(\delta_0)$ and $p^2$ are presented in units of the energy of $KB^*$ threshold.}
		\label{bsavere}
	\end{figure}

	In Figures \ref{bsava} and \ref{bsavere}, we present the L\"uscher-based amplitude fit results for the finite-volume $BK^*$ scattering data, whereas in Table \ref{tab:bsudavfit}, we list the corresponding best fit parameter values. We assume the zero range approximation for the amplitude and a linear fit form in $a$ here again. From either figure, it is evident that at all the $M_{ps}$ values, the continuum extrapolated amplitudes are consistent with zero inverse scattering length, which implies the system is in a regime where it cannot host any bound poles. The ground state energies being close to and consistent with the threshold leads to large uncertainties in the associated $p~cot(\delta_0)$ values, limiting the fits to be dominantly determined effectively by fewer degrees of freedom. We observe that the continuum extrapolated $p~cot(\delta_0)$ are consistent with zero at all the $M_{ps}$ values studied. It can also be observed that with such large uncertainties in $p~cot(\delta_0)$, the estimates for continuum binding energies would also be severely smeared out with huge uncertainties. This can clearly be observed in the case of $M_{ps}\sim0.5$ GeV, where the band representing continuum extrapolated amplitude overlaps with the bound state constraint curve (orange/cyan curve) over a large region along the $x$-axis.  

	\begin{table}[h]
		\setlength{\tabcolsep}{0.9mm}{
			\centering
			\renewcommand\arraystretch{1.9} 
			\begin{tabular}{|c|c|c|c|}
				\hline
				$M_{ps}$ [GeV] & $\chi^2$/d.o.f & $A^{[0]}/E_{KB}$ &  $A^{[1]}/E_{KB}$ \\
				\hline
				0.5 & 1.47/2 & $-0.027(^{+28}_{-31})$ & $0.23(^{+32}_{-23})$\\
				0.6 & 0.001/2 & $-0.03(^{+4}_{-6})$ &  $0.40(^{+1.11}_{-0.43})$ \\
				0.7 & 0.71/2 & $-0.02(^{+7}_{-5})$ & $0.19(^{+61}_{-58})$\\
				\hline
		\end{tabular}}
		
		\caption{Same as in Table \ref{tab:bsudavfit}, but for the scalar $bs\bar u\bar d$ channel. The best fit parameter values presented in third and fourth columns are in units of the energy of $KB$ threshold.}
	\end{table}
	
\begin{figure}[htbp]
		\centering
		\includegraphics[width=\linewidth]{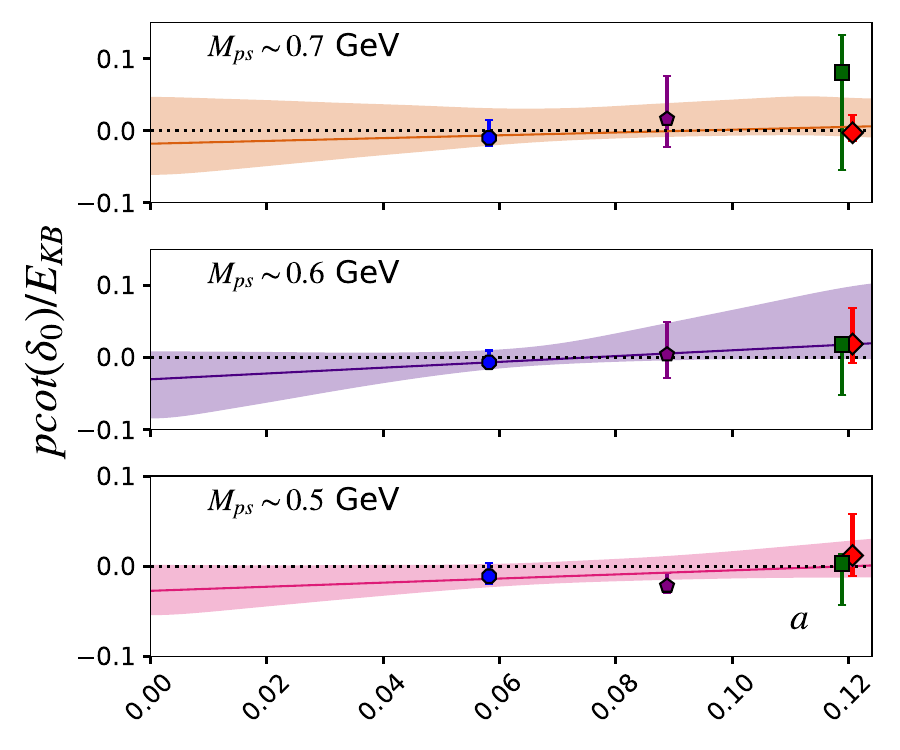}
		\caption{Same as in Figure \ref{bba}, but for the scalar channel in $KB$ scattering. $pcot(\delta_0)$ is presented in units of the energy of $KB$ threshold.}
		\label{bssca}
	\end{figure}

\begin{figure}[htbp!]
		\centering
		\includegraphics[width=\linewidth,height=8cm]{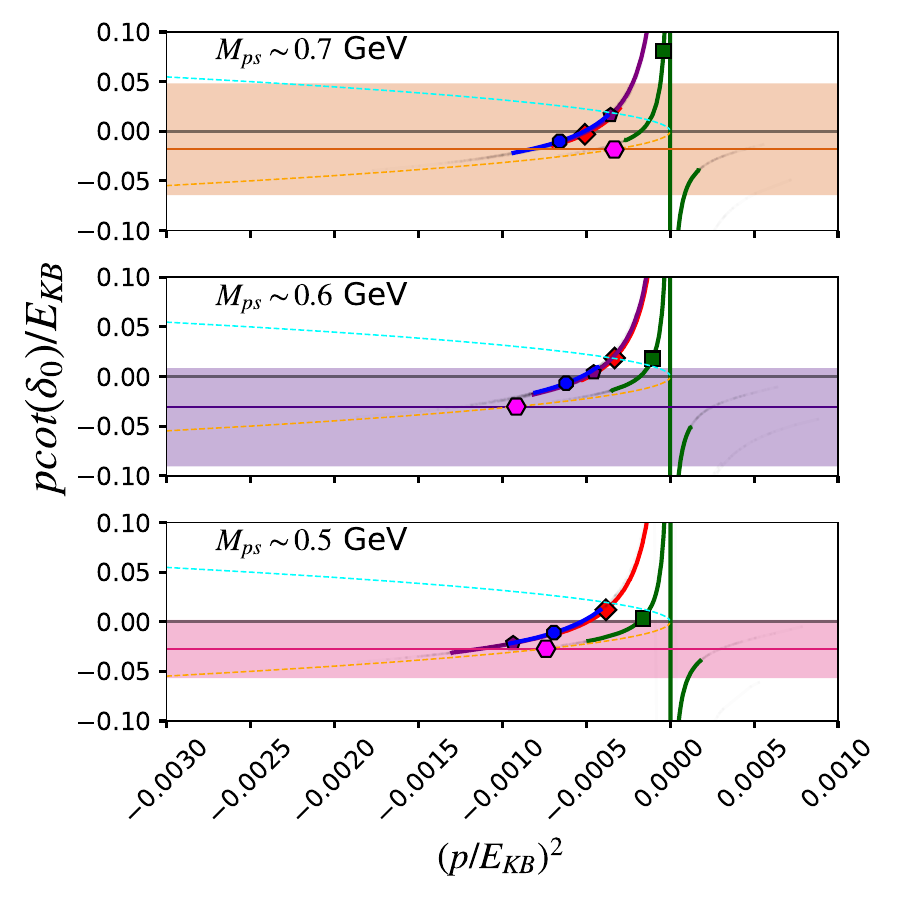}
		\caption{Same as in Figure \ref{bbere}, but for the scalar channel in $KB$ scattering. $pcot(\delta_0)$ and $p^2$ are presented in units of the energy of $KB$ threshold.}
		\label{bsscere}
	\end{figure}

	In Figures \ref{bssca} and \ref{bsscere}, we present the results for $BK$ scattering based on similar amplitude fits as discussed in the previous cases. Similar to the axialvector $bs\bar u\bar d$ tetraquarks, the continuum extrapolated amplitudes are consistently zero indicating infinite scattering length. The large uncertainties in the amplitudes also lead to large errors in the binding energies extracted from those. All of these are transparent enough to be readily inferred from Figures \ref{bssca} and \ref{bsscere}. In short, the $bs\bar u\bar d$ finite-volume data at hand does not support any statistically significant signatures for a bound axialvector or scalar $bs\bar u\bar d$ tetraquark. Similar to the arguments in stated in the case of $bb\bar u\bar d$, future lattice simulations closer to the physical pion mass $M_{\pi}$ are essential to extend our findings towards the chiral limit.
	
	\section{\label{sec:conclusion} Summary and Conclusions}
	We present a lattice QCD investigation of isoscalar tetraquark channels with one or more valence bottom quarks on lattice QCD ensembles generated by the MILC collaboration with $N_f=2+1+1$ dynamical HISQ fermions. The details of the lattice QCD ensembles used are listed in Table \ref{ensemb_table}. The valence quark dynamics are studied using an overlap formulation of the lattice fermion action for the quark masses up to the charm quark mass, whereas an NRQCD formulation was utilized to describe the bottom quark evolution. Utilizing four different ensembles varying in spatial volumes and lattice spacings involved, we account for potential discretization and finite-volume effects on the channels studied. 
	
	\begin{figure*}[htb!] 
		\centering
		\resizebox{\linewidth}{!}{\input{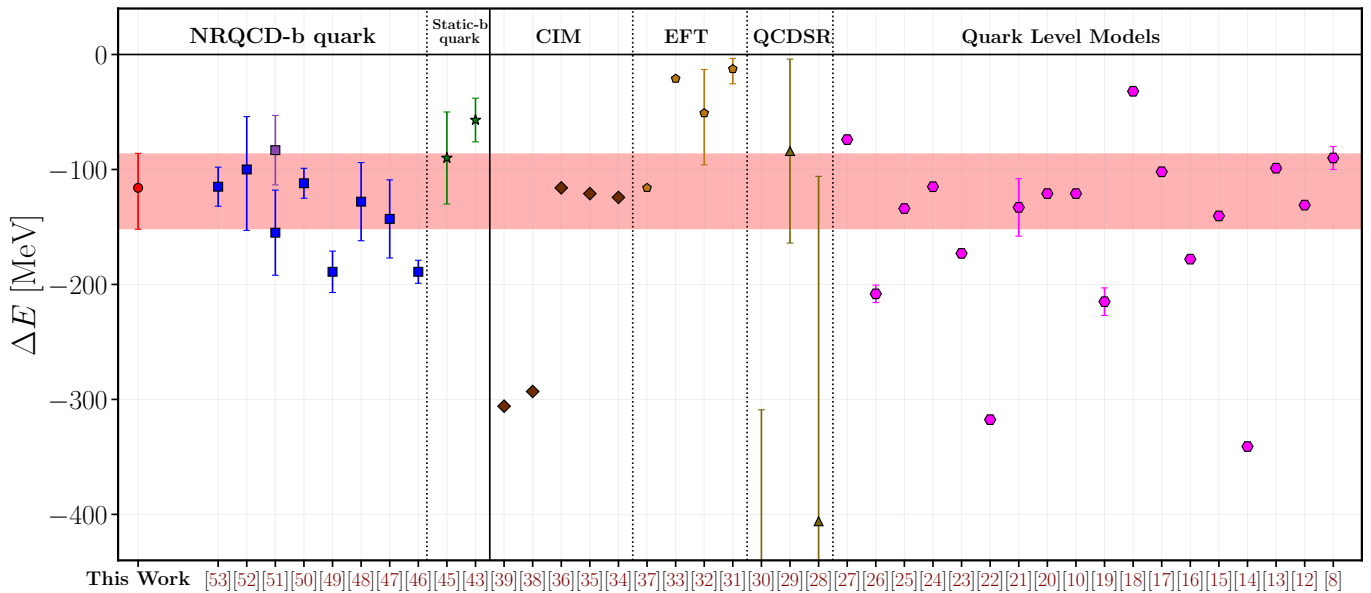}}
		\caption{Summary of binding energy predictions for the $bb\bar{u}\bar{d}$ system as made in various calculations. Our estimate is shown on the extreme left with a red marker in the left-most pane. A darker vertical line separates lattice determinations from phenomenological predictions. The first vertical panel from the left contains results from previous lattice determinations treating bottom quark dynamics with the NRQCD formalism (represented with blue markers). Two markers associated with Ref. \cite{Aoki:2023nzp} correspond to the estimates from an elastic analysis (blue) and a coupled $BB^*$-$B^*B^*$ analysis (violet). The second left panel contains lattice estimates assuming a static bottom quark (green markers). Phenomenological predictions are categorized as Chromomagnetic Interaction Models (CIM), Effective Field Theory (EFT), QCD Sum Rules (QCDSR), and Quark-Level Models, represented by brown, orange, yellow, and magenta markers, respectively. The $x$-axis tick labels indicate the references and can be identified with the vertical grids.}
		\label{fg:bb_summary}
	\end{figure*}
	
	This study addresses three isoscalar tetraquark channels involving bottom quarks, by investigating the respective finite-volume ground state energies. The first one is the isoscalar axialvector $bb\bar u\bar d$ channel, in which we investigate the low energy finite-volume eigenenergies across the ensembles, followed by a finite-volume scattering analysis, \ala~L\"uscher, to estimate the binding energy. Along the way, we also employ %recently proposed 
	box-sink smearing procedures to identify the ground state saturation in the correlators we utilize. This study of $bb\bar u\bar d$ tetraquarks is an extension to our previous calculation \cite{Junnarkar:2018twb}, with significant advancements made with regard to a more appropriate determination of ground state plateau and the binding energy, and more importantly, a finite-volume analysis. The two remaining channels are isoscalar $bs\bar u\bar d$ tetraquarks with axialvector and scalar quantum numbers, which are in continuation of the spectroscopy program we have performed in the bottom charm sector \cite{Padmanath:2023rdu, Radhakrishnan:2024ihu}. 
	
	In all the three channels we have studied, we have utilized operator bases composed of local two-meson interpolators corresponding to the lowest lying two-meson thresholds in the respective channels and local diquark-antidiquark interpolators and extracted the ground state energy. Finite volume ground state eigenenergies in the $bb\bar u\bar d$ sector indicate negative energy shifts with respect to the $BB^*$ threshold, suggesting the possibility of a bound state $T_{bb}$ tetraquark, with binding energy $\Delta E_{T_{bb}}(1^+)=-116(^{+30}_{-36})$ MeV. The binding energy estimate is based on a finite-volume analysis also accommodating for a lattice spacing dependence and is found to be insensitive to the light pseudoscalar meson mass $M_{ps}$ involved. The ground state finite-volume eigenenergies in the $bs\bar u\bar d$ sector do not indicate any statistically significant evidence for a bound state with either axialvector or scalar quantum numbers. 
	
    In Tables \ref{tab:pheno_tbb} and \ref{tb:latt_tbb} and Figure \ref{fg:bb_summary}, we present a compilation of various estimates for the binding energy of $T_{bb}$ tetraquarks following lattice as well as nonlattice methodologies, together with the estimate we arrive from this investigation. We observe our new estimate, that involved a conservative determination of the ground state energies followed by a finite-volume scattering analysis, is consistent with all the predictions from other lattice QCD determinations, except two outliers with deeper binding. Owing to the presence of heavy quarks, the finite-volume binding energy estimates in this heavy tetraquark system could be affected by discretization effects. Ours is the only calculation that has utilized finite-volume data at three lattice spacings that go as fine as $\sim0.06$ fm, and also attempt a finite-volume analysis in the $T_{bb}$ system. Our work utilizes a chiral quark action for the light, strange and charm quark fields in contrast to most other existing studies, which commonly uses a Wilson-clover action, offers a complementary perspective. Moreover, using the same fermion action across a range of systems provides a valuable consistency check on the binding energies in these tetraquark systems. 
    
    Despite the relatively heavy pseudoscalar meson masses $M_{ps}$ used and the lack of meson-meson operators with separately momentum-projected mesons in this study, our estimate seem to agree very well with existing lattice determinations that include them. Although this is encouraging, future investigations with $M_{ps}$ values closer to the physical point, incorporating local as well as bilocal meson-meson interpolators, accompanied by  a rigorous finite-volume analysis, are essential to reach at rigorous and conservative estimates on the bindings also accommodating for various unaccounted systematics in the systems addressed here. 
    
\begin{figure}[htb!] 
		\centering
		\resizebox{\linewidth}{!}{\input{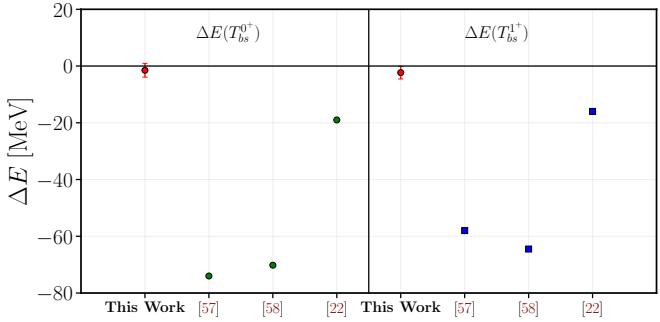}}
		\caption{Summary of binding energy predictions for the $bs\bar{u}\bar{d}$ system from various calculations, presented separately for the scalar and axialvector channels, divided by a vertical line. Our value presented in red markers corresponds to the finite-volume energy splitting observed in the largest volume ensemble at the lowest pseudoscalar mass studied. }
		\label{fg:bs_summary}
	\end{figure}

    No lattice determination, including this work, unambiguously suggests the existence of $bs\bar u\bar d$ bound state, whereas a few phenomenological approaches argue possible binding. We collect existing predictions for this system in Figure \ref{fg:bs_summary} and Table \ref{bsud_table}. We remark that there is only one other lattice investigation of the $bs\bar u\bar d$ system \cite{Hudspith:2020tdf}, where the author claims no evidence for any bound $bs\bar u\bar d$ state. Since there is no estimate for the ground state energy provided in Ref. \cite{Hudspith:2020tdf} for the $bs\bar u\bar d$ system, and hence is missing in Figure \ref{fg:bs_summary}. For similar reasons, as stated in the above paragraph, future lattice simulations closer to the physical point and with interpolating operators that mimic bilocal meson-meson structures are crucial in assessing the interactions towards the chiral limit and to reaffirm or to rule out our findings.
	
	In summary, our lattice setup suggests a deeply bound tetraquark in the axialvector $bb\bar u\bar d$ system with binding energy $\mathcal{O}(100$ MeV$)$, whereas an unbound $bs\bar u\bar d$ system in both axialvector and scalar quantum channels. Note that a relatively shallow yet deeply bound state with axial-vector and scalar quantum numbers for the $bc\bar u\bar d$ system, with a binding energy of approximately 40 MeV, was recently reported using the same lattice setup employed here \cite{Padmanath:2023rdu, Radhakrishnan:2024ihu}. This result helps bridge the sequence of these tetraquarks, which have valence quark contents ranging from bottom to strange. While our inferences on the $bb\bar u\bar d$ system and the $bs\bar u\bar d$ system turn out to be consistent with other existing lattice determinations \cfex~Ref.~\cite{Leskovec:2019ioa}, the disagreements in binding energies for the $bc\bar u\bar d$ system with other existing lattice determination \cite{Alexandrou:2023cqg} remains to be understood, particularly considering the effect of discretization in these heavy hadrons.

	\begin{table*}[htp]
		\setlength{\tabcolsep}{0.5mm}{
			\centering
			\renewcommand\arraystretch{1.2} 
			\begin{tabular}{|c|c|c|c|c|c|c|c|c|c|}
				\hline 
				Ref. [Year)]     &$N_{f}$           &$a~[fm]$         & $n_a$ &$m^{sea}_{ps}$ [MeV]    &$S_{q}^{sea}$       &$S_{q}^{val}$  &$S_{b}^{val}$     & Amplitude  &$\Delta E_{T_{bb}}(1^{+})$   
				\\
				&          &        &  & &    & &&Analysis  & [MeV]
				\\
				
				\hline
				\cite{Bicudo:2012qt} (2012)     & 2  &  0.079    & 1& 340     & Twisted mass    &Twisted mass &Static    & \textbf{Potential} & $-57(19)$     \\
				\cite{Bicudo:2015vta} (2015)     &2   &0.042,0.079  &  2  & 340, 352    & Twisted mass    &Twisted mass   &Static  & \textbf{Potential} &-90(40)     \\
				\cite{Francis:2016hui} (2017)     & 2+1 &   0.0899  &1 &  164-415  &  Wilson-clover  & Wilson-clover & NRQCD  &  No&-189(10)  \\
				\cite{Junnarkar:2018twb} (2019)    &2+1+1     &  0.0582-0.1207& \textbf{3}   &153-689    &HISQ           &Overlap &  NRQCD &  No  &-143(34)  \\
				\cite{Leskovec:2019ioa} (2019) &  2+1     &  0.0828-0.1141 & 3 &  139-431  &   Domain wall      &  Domain-wall   &  NRQCD & \textbf{L\"uscher-based}
				&-128(34)   \\
				\cite{Mohanta:2020eed} (2020) & 2+1    &  0.09-0.15& 3 & - & Asqtad  &   HISQ        &  NRQCD & No &-189(18)   \\
				\cite{Hudspith:2023loy} (2023) &  2+1    & 0.085& 1 & 220-420  & Wilson-Clover &   Wilson-Clover     &  NRQCD & No & -112(13)     \\
				\cite{Aoki:2023nzp} (2023) &  2+1    & 0.0907& 1 & 416-701  & Wilson-Clover &  Wilson-Clover     &  NRQCD & \textbf{HALQCD}  & -155(17)(20)     \\
				& & & & & & & & & -83.0(10.2)(20)\\
				\cite{Alexandrou:2024iwi} (2024) &  2+1+1    & 0.0872-0.1510& \textbf{3} & 217-313 & HISQ  &   Wilson-clover       & NRQCD & No & $-100(10)(^{+36}_{-43})$   \\
				\cite{Colquhoun:2024jzh} (2024) &  2+1    & 0.08& 1 & 164-707 & Wilson-Clover  &   Wilson-clover       & NRQCD & No  & $-115(17)$   \\
				\hline
				This work &  2+1+1    & 0.0582-0.1207& \textbf{3} & 217-319 & HISQ  &   Overlap       & NRQCD & \textbf{L\"uscher-based}
				& $-116(^{+30}_{-36})$   \\
				\hline 
		\end{tabular}}
		\caption{Binding energy $\Delta E_{T_{bb}}(1^{+})$ predictions from various lattice QCD studies. The results are presented along with some of the relevant technical details pertaining to the respective lattice investigations. $n_a$ refers to the number of distinct lattice spacings used in each calculation and the bold faced numbers are cases, where a continuum extrapolation was performed. Two numbers associated with Ref. \cite{Aoki:2023nzp} correspond to the estimates from an elastic analysis (top) and a coupled $BB^*$-$B^*B^*$ analysis (bottom). }
		\label{tb:latt_tbb}
	\end{table*}
	\begin{table}[htp]
\setlength{\tabcolsep}{0.5mm}{
\centering
\renewcommand\arraystretch{1.2} 
\begin{tabular}{|c|c|c|}
\hline 
Ref. (Year)   & $\Delta E_{T_{bb}}(1^{+})$ (MeV) & Table No (page) \\
\hline
\multicolumn{3}{|c|}{\textbf{Quark-Level Models}} \\
\hline
\cite{Carlson:1987hh}(1988)      & -90(10) & Table \rom{5}  \\
\cite{Silvestre-Brac:1993zem}(1993) &  -131 & Table \rom{3}  \\
\cite{Brink:1998as}(1998)       &  -98.9 & Table \rom{4}  \\
\cite{Vijande:2003ki}(2004)     &  -341 & Table \rom{3} \\
\cite{Janc:2004qn}(2004)        &  -140.2 & Table \rom{1}  \\
\cite{Vijande:2006jf}(2006)     &  -178 & Table \rom{3}  \\
\cite{Ebert:2007rn}(2007)        & -102 & Table \rom{3}  \\
\cite{Zhang:2007mu}(2008)        & -32 & Table \rom{4} \\
\cite{Karliner:2017qjm}(2017)    & -215(12) & Table \rom{1}  \\
\cite{Eichten:2017ffp}(2017)     & -121 & Table \rom{2}  \\
\cite{Park:2018wjk}(2019)        & -121 &  inline (10)\\
\cite{Braaten:2020nwp}(2021)     & -133(25) & Table \rom{6} \\
\cite{Tan:2020ldi}(2020)         & -317.6 & inline (7) \\
\cite{Meng:2021agn}(2021)        & -173 & Table \rom{2} \\
\cite{Kim:2022mpa}(2022)         & -115 & inline (14) \\
\cite{Praszalowicz:2022sqx}(2022)  & -134 & Table \rom{4} \\
\cite{Wu:2022gie}(2023)          & -208.2(7.6) & Table \rom{3} \\
\cite{Song:2023izj}(2023)        & -74 & Table \rom{7} \\
\cite{Meng:2023jqk}(2023)        & -151.6 & Table \rom{3} \\
\hline
\multicolumn{3}{|c|}{\textbf{QCD Sum Rules}} \\
\hline
\cite{Navarra:2007yw}(2007)      & -406(300) & inline (5) \\
\cite{Wang:2017uld}(2018)        & -84(80) & Table \rom{4} \\
\cite{Agaev:2018khe}(2019)      & -569(260)& inline (5)\\
\hline
\multicolumn{3}{|c|}{\textbf{Effective Field Theory}} \\
\hline
\cite{Wang:2018atz}(2019)        & $-12.6(^{+9.2}_{-12.9})$ & Table \rom{3} \\
\cite{Liu:2019stu}(2019)         & $-51(^{38}_{45})$ & Table \rom{4} \\
\cite{Dai:2022ulk}(2022)         & -21 & Table \rom{5} \\
\cite{Brambilla:2024thx}(2024) & -116 & inline (4) \\
\hline
\multicolumn{3}{|c|}{\textbf{Chromomagnetic Interaction Models}} \\
\hline
\cite{Lee:2009rt}(2009)          & -124.3 & Table \rom{4} \\
\cite{Luo:2017eub}(2017)         & -121 & Table \rom{6}\\
\cite{Cheng:2020wxa}(2021)       & -116 & Table \rom{6}\\
\cite{Guo:2021yws}(2022)         & -293.1 & Table \rom{4} \\
\cite{Liu:2023vrk}(2023)         & -305.9 & Table \rom{8} \\
\hline
\end{tabular}}
\caption{Comparison of $\Delta E_{T_{bb}}(1^{+})$ results from various phenomenological studies, categorized by approach. The third column is a pointer to the table within the corresponding reference, from where the number has been taken. Those cases where numbers are provided inline, are indicated so along with a number referring to the page number in the published version of the article, where this estimate is quoted.}
\label{tab:pheno_tbb}
\end{table}
	
	\begin{table}[h]
		\setlength{\tabcolsep}{0.9mm}{
			\centering
			\renewcommand\arraystretch{1.9} 
			\begin{tabular}{|c|c|c|c|}
				\hline
				Ref (Year)& Approach &$\Delta E({T^{0^+}_{bs}})$ & $\Delta E({T^{1^+}_{bs}})$ \\
				\hline
				\cite{Huang:2019otd} (2019)& Color decolorization Model & -74 & -58\\
				\cite{Chen:2018hts} (2018)& SU(3) chiral quark Model &-70.2 & -68 \\
				\cite{Tan:2020ldi} (2020) & Alternate quark Model &-19 & -16 \\
                    \cite{Chen:2023syh} (2020) & Quark potential Model &-3 & -1 \\
				\cite{Agaev:2019wkk} (2019)& QCD sum rule &-394(170) & -\\
				\cite{Zouzou:1986qh} (1986)& non-chiral Model &- & unbound\\
				\hline
		\end{tabular}}
		\caption{Summary of previous phenomenological calculations of \( T_{bs} \), listing the reference, approach, and results for the scalar and axial-vector cases.}
		\label{bsud_table}
	\end{table}
	%%%%%%%%%%%%%%%%%%%%%%%%%%%%%%%%%%%%%%%%%%%%%%%%%%%%%%%%%%%%%%%%%%%%%%%%%%%%%%%%%%%%%%%%%%%%%%%%
	%%%%%%%%%%%%%%%%%%%%%%%%%%%%%%%%%%%%%%%%%%%%%%%%%%%%%%%%%%%%%%%%%%%%%%%%%%%%%%%%%%%%%%%%%%%%%%%%
	%%%%%%%%%%%%%%%%%%%%%%%%%%%%%%%%%%%%%%%%%%%%%%%%%%%%%%%%%%%%%%%%%%%%%%%%%%%%%%%%%%%%%%%%%%%%%%%%
	\vspace{0.1cm}
	\section{Acknowledgments}
	The authors would like to thank Navdeep Singh Dhindsa and Tanishk Shrimal for valuable discussions. This work is supported by the Department of Atomic Energy, Government of India, under Project Identification Number RTI 4002. Computations were carried out on the Cray-XC30 of ILGTI, TIFR (which has recently been closed), and the computing clusters at the Department of Theoretical Physics, TIFR, Mumbai, and IMSc Chennai. We are thankful to the MILC collaboration and, in particular, to S. Gottlieb for providing us with the HISQ lattice ensembles. We would also like to thank  Ajay Salve, Kapil Ghadiali, G. Srinivasan, and T. Chandramohan for computational support. We thank the authors of Ref. \cite{Morningstar:2017spu} for making the~{\it TwoHadronsInBox} package public. MP gratefully acknowledges support from the Department of Science and Technology, India, SERB Start-up Research Grant No. SRG/2023/001235.
	\bibliography{references}

\end{document}